\documentclass{jfm}
\usepackage{epsfig,psfrag}
\usepackage{amsmath}
\usepackage{amsfonts}
\usepackage{mathrsfs}
\usepackage{amssymb}
\usepackage{bm}
\usepackage{color}
\usepackage{natbib}
\usepackage{mdwlist,url}
\usepackage{xr,rotating}

\begin{document}

\newcommand\beq{\begin{equation}}
\newcommand\eeq{\end{equation}}
\newcommand\beqa{\begin{eqnarray}}
\newcommand\eeqa{\end{eqnarray}}
\newcommand{\Sy}{{\cal S}}
\newcommand{\U}{{\cal U}}
\newcommand{\K}{{\cal K}}
\newcommand{\T}{\mathsfi {T}} 
\newcommand{\bx}{\mathbf{x}}
\newcommand{\by}{\mathbf{y}}
\newcommand{\bz}{\mathbf{z}}
\newcommand{\hx}{\mathbf{ \hat{x}}}
\newcommand{\hy}{\mathbf{ \hat{y}}}
\newcommand{\hz}{\mathbf{ \hat{z}}}
\newcommand{\hg}{\mathbf{ \hat{g}}}

\def\half{\frac{1}{2}}
\def\quart{\frac{1}{4}}
\def\ud{{\rm d}}
\def\eps{\epsilon}

\title[Plane Couette flow with spanwise stratification]{Layer formation and relaminarisation in plane Couette flow with spanwise stratification}

\author{Dan Lucas\aff{1}\corresp{\email{d.lucas1@keele.ac.uk}} C. P. Caulfield\aff{2,3} \& Rich R.  Kerswell\aff{3} }
\affiliation{\aff{1}{School of Computing and Mathematics, Keele University, Staffordshire, ST5 5BG}
\aff{2}{BP Institute, University of Cambridge, Madingley Rise, Madingley Road, Cambridge, CB3 0EZ, UK}
\aff{3}{Department of Applied Mathematics \& Theoretical Physics, University of Cambridge, Centre for Mathematical Sciences, Wilberforce Road, Cambridge, CB3 0WA, UK}}
\maketitle
\begin{abstract}

 In this paper we investigate the effect of stable stratification on plane Couette flow when gravity is oriented in the spanwise direction. When the flow is turbulent we observe near-wall layering and associated new mean flows in the form of large scale spanwise-flattened streamwise rolls. The layers exhibit the expected buoyancy scaling  $l_z\sim U/N$  where $U$ is a typical horizontal velocity scale and $N$ the buoyancy frequency. We associate the new coherent structures with a stratified modification of the well-known large scale secondary circulation in plane Couette flow. We find that the possibility of the transition to sustained turbulence is controlled by the relative size of this buoyancy scale to the spanwise spacing of the streaks. {In parts of parameter space transition can also be initiated by a newly discovered linear instability} in this system (Facchini et. al. 2018 \emph{J. Fluid Mech.} vol. 853, pp. 205-234). When wall-turbulence can be sustained the linear instability opens up new routes in phase space for infinitesimal disturbances to initiate turbulence. When the buoyancy scale suppresses turbulence the linear instability leads to more ordered nonlinear states, with the possibility for intermittent bursts of secondary shear instability. 
 \end{abstract}
\section{Introduction}

When turbulence occurs in the presence of a stably-stratified background gradient of density, it is common to observe the spontaneous formation of layers of well-mixed fluid separated by relatively sharp gradients or interfaces of density \citep{Park1994, Holford:1999uc, Holford:1999,Oglethorpe:2013cv, Thorpe:2016ga,2016JPO....46.1023F,Leclercq:2214839}. Such behaviour is of direct relevance to the atmosphere, oceans and various other environmental and industrial flows as these interfaces act as barriers to mixing and transport, and have a significant effect on the overall energetics of the flow. 

Stratification also has a tendency to suppress vertical velocities due to the restoring force of gravity, thereby creating an inherent anisotropy in the flow when the stratification is large. Scaling arguments by \cite{Billant:2001cs} predict that such anisotropy occurs with a vertical `buoyancy' length scale $l_z\sim U/N$ where $U$ is a typical horizontal velocity scale and $N$ is the buoyancy or Brunt-V\"ais\"al\"a frequency. This scaling is observed  ubiquitously in stratified flows exhibiting layers,
{although relatively few}
well-defined physical mechanisms are able to make direct contact with it.

The motivation of this work is primarily to increase the connectivity between various  approaches to the above problem of layer formation. The recent results presented in \cite{LCK} (hereafter LCK) have demonstrated{, using a triply-periodic domain,} that nonlinear exact coherent structures associated with layer formation can be traced through parameter space and connected to linear instabilities of a horizontally varying basic flow. 
Here we will consider a more physically realistic flow which has a base flow with horizontal shear. The flow is forced naturally by  moving boundaries in a plane Couette flow (pCf) system with spanwise stratification (hereafter referred to as HSPC {for horizontal stratified plane Couette}).

This paper constitutes, along with the recent paper by \cite{facchini_2018}, the first exploration of the effect of spanwise stratification on plane Couette flow dynamics. 
{Mainly considering flows at high Prandtl}
number, \cite{facchini_2018} have shown that for the spanwise stratified version of plane Couette flow,  a new linear instability appears  when the geometry permits resonances between internal gravity waves. 
{However, the instability only arises} 
in geometries which allow the {required} critical relationship between vertical and streamwise wavelengths, given by the {appropriately} Doppler-shifted dispersion relationship. Therefore in many cases, and in particular at {sufficiently} low stratifications, a subcritical transition scenario comparable to unstratified pCf is observed.
{These two {possible} routes to turbulence make this flow geometry a particularly attractive one to investigate further the mechanisms by
which stratification can affect turbulent flows, as well as identifying whether the initial (subcritical or supercritical) transition mechanisms leave a qualitatively identifiable imprint on the ensuing turbulent dynamics.}

There is an increasing body of literature considering the similar configuration for  (axially) stratified Taylor-Couette flow \citep{Molemaker:2001hc,Shalybkov:2005,Oglethorpe:2013cv, Leclerq_2016,Leclercq:2214839,Park_2017,Park_2018} from experimental, numerical and theoretical approaches. The primary focus for this flow has been the interplay between various instability mechanisms, in particular between the so-called centrifugal and stratorotational instabilities \citep{Leclerq_2016,Shalybkov:2005,Molemaker:2001hc}. In certain situations it is possible to observe the formation of layers, often confined near the walls, the source of which is the subject of continued debate. Part of the motivation of this work is to remove the effects of rotation and curvature from this system {by taking the narrow-gap limit to try to uncover the  universal, persistent features of such flows. }

Also of interest is the comparison to the recently studied situation where the shear and stratification gradients are aligned. This case has been examined to study  spatiotemporal dynamics,
{in particular the extent to which stratification
can suppress turbulence and lead to relaminarisation} \citep{Deusebio:2015, Taylor_2016}, irreversible mixing and layer robustness properties \citep{Zhou:2017dd,zhou_2017} and the effect of stratification on underlying exact coherent structures \citep{Clever_1992, Clever:2000, olvera_2017,Deguchi_2017}.

In this paper we establish that, as in the vertically sheared case, only {relatively} weak stratification  {(in a sense which we make precise below) leads to suppression of} the subcritical turbulence present in plane Couette flow, 
{at least in flow geometries of aspect ratio $O(1-10)$.}
{Perhaps unsurprisingly, when the flow is linearly unstable
to the instability identified by \cite{facchini_2018}, yet is still {below} the relaminarisation boundary,  turbulent transition occurs supercritically. By considering flow geometries at similar parameters that can either admit or not this linear instability, we  
conclude that the supercritically-triggered turbulent state is similar
to the subcritically-triggered turbulent state, {namely having the characteristics of the self-sustaining process or wave-vortex interaction (SSP/VWI) mechanism \citep{Hall:1991eq,Waleffe:1997ia,HALL:2010cn}.}
Furthermore, when the flow is linearly unstable, yet above the relaminarisation boundary, although perturbations saturate at finite amplitude, they remain {relatively} ordered, with only highly spatio-temporally intermittent `bursts' of disordered motions.
}

{ In general, we }
observe
{that}
stratification {always appears to have an effect on the dynamics,}
by altering the large scale secondary flow patterns. These 
{modified} mean flows 
{induce} the formation of density layers near the walls which exhibit the familiar $U/N$ {buoyancy} scaling, 
{although the physical mechanism leading to their formation is different from that underlying the previously identified `zig-zag'-like instabilities}
(\cite{Billant:2000jg}, LCK). At fixed Reynolds number, we demonstrate that the relaminarisation boundary as 
{the stratification becomes relatively stronger}
corresponds to the intersection of this buoyancy scale of the secondary flow and the  length scale associated with the spacing of the streaks. This 
{implies} that {subcritical} turbulence transition is controlled by a different competition of length scales compared to the {wall-normal stratified} case where \cite{Deusebio:2015} demonstrated that the competition between the Monin-Obukhov length scale and the (inner) viscous scale controls relaminarisation and intermittency. 

{Interestingly, we find that for sufficiently large Reynolds numbers, the 
critical strength of stratification leading to relaminarisation 
is higher in the spanwise-stratified case considered here than in the wall-normal stratified case of \cite{Deusebio:2015}.}
{This observation of turbulence `surviving' at higher stratification is consistent with the commonly observed phenomenon that shear flows where the velocity gradient is orthogonal to the density gradient (usually referred to as `horizontal shear') allow for stronger injection of perturbation kinetic energy from this shear into ensuing turbulent flows \citep{Jacobitz:1998di}, than in flows with `vertical shear' where the velocity and density gradients are parallel.}
{Furthermore,  although at higher Reynolds number in some appropriate flow geometries, turbulence can be triggered supercritically  via the linear instability identified by \cite{facchini_2018}, rather than subcritically,  the ultimate {sustained} turbulent attractor is, as expected, the same.}

{To illustrate these various issues and observations, the rest of this} paper is organised as follows. Section 2 provides the formulation of the numerical simulation and defines the various parameters of interest. {Section 3 presents the results of the numerical simulations conducted when the base flow is linearly stable, {in particular discussing}
the formation of layers and their influence on relaminarisation. Section 4 considers linearly unstable flows and how the relaminarisation boundary 
{affects} 
the nonlinear dynamics. Finally, section 5 provides some {further}
discussion and conclusions.}

\section{Formulation}\label{sec:form}

We begin by considering the following version of the wall-forced, incompressible, Boussinesq equations

\begin{align}
\frac{\partial \bm u^*}{\partial t^*} + \bm u^*\cdot\nabla^*\bm u^*
 +\frac{1}{\rho_0}\nabla^*p^* &= \nu \Delta^* \bm u^*  - \frac{\rho^*g}{\rho_0}\hz, \\ 
\frac{\partial \rho^*}{\partial t^*}+\bm u^*\cdot\nabla^*\rho^* +\bm u^*\cdot\nabla^*\rho_B&=  \kappa \Delta^* \rho^*\\
\nabla^*\cdot \bm u^* & =0 \\
\end{align}
where $\bm u^*(x,y,z,t) = u^* \hx+v^* \hy+w^* \hz$ is the three-dimensional
velocity field, $\nu$ is the kinematic viscosity, $\kappa$ is the molecular diffusivity, $p^*$ is the pressure, $\rho_0$ is an appropriate reference density and $\rho^*(x,y,z,t)$ the varying part of the density away from the background linear density profile $\rho_B=-\beta z,$ i.e. $\rho_{\mathrm{total}}=\rho_0+\rho_B(z)+ \rho^*(x,y,z,t)$. We impose periodic boundary conditions in the streamwise, $x,$ and spanwise, $z,$ directions with no slip walls for $\bm{u}^*$ and no flux for $\rho^*$ in $y:$ 
\begin{align}
\bm{u}^*(x,y=-L_y,z)=(-U_y,0,0) &\qquad \bm{u}^*(x,y=L_y,z)=(U_y,0,0) \\ \left.\frac{\partial \rho^*}{\partial y}\right|_{y=-L_y}& = \left.\frac{\partial \rho^*}{\partial y}\right |_{y=L_y}=0
\end{align}
 and consider domains
$(x,y,z) \in [0,L_x]\times[-L_y,L_y]\times[0,L_z].$
{Within this coordinate system, gravity may be thought of as pointing in the (negative) vertical direction, and the pCf induces horizontal shear, hence we refer to the flow as HSPC.}

The system is naturally non-dimensionalised using the characteristic length scale $L_y$, characteristic time scale $U_y/L_y$ and density gradient scale $\beta=-\nabla \rho_B\cdot \hz$  to give
\begin{align}
\frac{\partial \bm u}{\partial t} + \bm u\cdot\nabla\bm u +\nabla p 
&= \frac{1}{Re} \Delta \bm u - F_h^{-2} \rho \hz\label{NSu},\\ 
\frac{\partial \rho}{\partial t} + \bm u\cdot\nabla\rho &= w + \frac{1}{Re Pr}\Delta \rho\\
\nabla\cdot \bm u &=0
\end{align}
where we define the Reynolds number $Re$, 
{an appropriate background} 
horizontal Froude number $F_h$, the Prandtl number $Pr$ and buoyancy frequency $N$ as
\begin{equation}
Re := \frac{U_yL_y}{\nu}, \quad F_h:= \frac{U_y}{N L_y}
, \qquad Pr := \frac{\nu}{\kappa},\qquad N^2:=\frac{g\beta}{\rho_0} .\label{eq:params}
\end{equation}

{We choose to use the (horizontal) Froude number $F_h$ as the appropriate measure of the relative strength of the background shear to the background stratification as this is exactly in the same form as the Froude number considered by \cite{facchini_2018}. For the flow considered by \cite{Deusebio:2015}, where the shear and density gradient are parallel, the natural parameter is the bulk Richardson number $Ri$, which is 
mathematically equivalent, within this formulation, to the inverse square of the Froude number, i.e. $ Ri \equiv F_h^{-2}$.  It must always be remembered that for the flow we are considering here, the stratification does not act directly against wall-normal motions, and so 
the conventional interpretation 
of a Richardson number (or equivalently the inverse 
square of a Froude number) as quantifying
the relative significance of potential energy to kinetic energy in the background shear must be done with care, and in particular there is no {\it a priori} reason why large values of $Ri$ (or small values of $F_h$) will lead to the flow being stable. However, it is reasonable to draw some analogies between  the flow considered here and the flow considered in \cite{Deusebio:2015}, as their parameter $Ri$ and our parameter $F_h$ are both ratios of 
the characteristic time scales associated with the background density and velocity distributions. Furthermore, as is apparent from the governing equations, it is the inverse square of the Froude number which quantifies the coupling between the density and velocity fields.}

To characterise the basic energetics of the flows we define
\begin{align} 
\mathcal{K} :=\frac{1}{2}\langle|\bm u'|^2\rangle_V,&\qquad \epsilon := \frac{1}{Re}\left\langle\left| \nabla \bm u'\right|^2 \right \rangle_V, \qquad\\
\mathcal{\chi} := \frac{1}{PrRe}\langle|\nabla \rho|^2\rangle_V,& \qquad u_\tau:=\sqrt{\tau_w/\rho_0}; \quad \tau_w:=\left . \nu \frac{\partial u}{\partial y} \right |_{y=\pm1}, 
\label{eq:diag}
\end{align}
where $\bm u'=\bm u - \langle \bm u\rangle_x$ is the fluctuation velocity, $\mathcal{K}$ is the turbulent kinetic energy, $\epsilon$ is the turbulent dissipation rate, $\chi$ is the dissipation rate of density variance and $u_\tau$ is the friction velocity {defined in terms of} the wall shear stress $\tau_w$.  {In these definitions, } $\langle (\cdot)  \rangle_V:= \int \! \! \int \! \! \int (\cdot)\,dxdydz/(2L_xL_z)$ denotes a volume average,  $\langle (\cdot)  \rangle_x:= \int (\cdot)\,dx/L_x$ denotes a streamwise average 
$\langle{\cdot}\rangle_t = [\int_0^T (\cdot) \ud t]/T$ {denotes a time average,} where $T$ is normally {close to} the full simulation time {(having removed the initial transient spin-up from the initial condition)}.  

We fix $Pr=1$ throughout and solve the equations numerically using the DIABLO direct numerical simulation (DNS) code \citep{Taylor2008} which {is} mixed pseudospectral in $(x,z)$ with {second order} finite difference in the wall-normal direction and {uses} {fourth order} Runge-Kutta ({for the} nonlinear \& buoyancy terms) and Crank-Nicolson ({for the} diffusion {terms}) timestepping. {The resolution $(N_x,N_y,N_z)$ is defined such that $N_x$ and $N_z$ are the number of Fourier collocation points in each direction and $N_y$ the number of finite difference points which are clustered near the wall.} Following \cite{moin_1998} and \cite{Deusebio:2015}, we maintain spatial convergence by ensuring that $\Delta x^+ \lesssim 8, \, \Delta z^+ \lesssim 4$ and $y_{10}^+\lesssim 10$ where $y^+_{10}$ is the tenth point from the wall and the `$+$' superscript represents the usual viscous scaling $l^+=u_\tau l/\nu.$  {The simulations to follow are initialised with a broad band uniform spectrum of perturbations having randomised phases with large enough amplitude ({10-20\%}
of the wall velocity) to trigger {(subcritically)} sustained turbulence.}

\begin{centering}
\begin{figure}
\includegraphics[width=\textwidth]{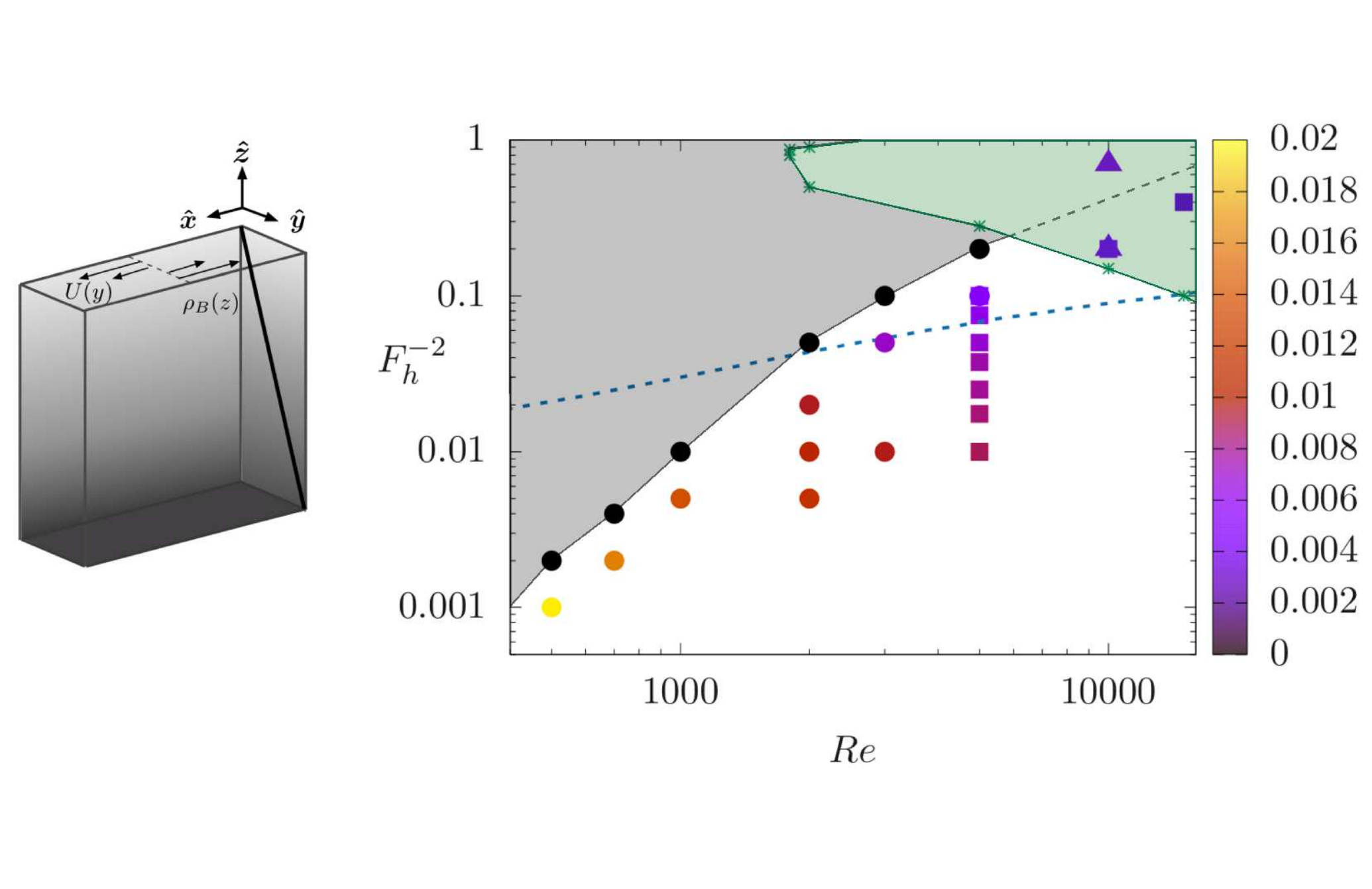}
\caption{\label{fig_TKE} Left: Schematic showing the base flow and the background stratification for HSPC. Right: ($Re,$ $F_h^{-2}$) parameter space from the DNS in table \ref{tab:DNS} coloured with turbulent kinetic energy $\langle{\mathcal{K}}\rangle_t$ showing the sustenance of the turbulence, where  black dots represent $\langle{\mathcal{K}}\rangle_t=0$ where the flow has relaminarised. Symbols corresponds to the groups of table \ref{tab:DNS}; circles have $L_x=4,$ squares have $L_x=\pi$ and triangles the domains permitting the linear instability as discussed in section \ref{sec:instab}. {The dashed blue line marks the theoretical model (using
Monin-Obukhov theory) 
which predicts the intermittency boundary well for wall-normal stratified pCf as reported in \cite{Deusebio:2015}. The bulk Richardson number $Ri$ in this model is equated to $F_h^{-2}$ for ease of comparison. The grey shaded region denotes the region of parameter space where the flow relaminarises, while the green region approximately denotes the region where the flow (for some combination of streamwise and vertical wavenumbers) is prone to the primary linear instability identified by \cite{facchini_2018}.}  {The dashed grey curve extends the relaminarisation boundary into the linearly unstable region, extrapolating the anticipated boundary now between linearly unstable, yet ordered dynamics, and sustained wall turbulence.} A triangle at $Re=10000,\, F_h^{-2}=0.7$ denotes the case where the flow is linearly unstable but the SSP/VWI mechanism is suppressed leading to qualitatively different (and non-turbulent) flows, as discussed in section \ref{sec:instab}.}
\end{figure}
\end{centering}


\section{Direct numerical simulations: subcritical turbulence}\label{sec:DNS}

We begin by presenting two sets of direct numerical simulations which were carried out to survey the $(Re,F_h)$ parameter space near 
{subcritical} transition to sustained turbulence, with different choices of streamwise domain length. Table \ref{tab:DNS} outlines the parameters and some single point statistics. {The right hand panel of figure} \ref{fig_TKE} shows $\langle{\mathcal{K}}\rangle_t$ in $(Re,\, F_h)$ space, indicating the laminar-turbulent boundary (note the relaminarised cases are not shown in table \ref{tab:DNS}). We find that a similar picture to the vertically shearing case of \cite{Deusebio:2015} emerges in that only {relatively weak} stratification is required to shift the critical Reynolds number {required} for sustained turbulence, and that stratification reduces the overall turbulent kinetic energy of the flow. Comparing specific values, we note that with horizontal mean shear,
 larger stratifications {for turbulent flows} are achievable than for the vertical case {for sufficiently high flow Reynolds numbers, as we also plot on the figure the curve associated with formula derived using Monin-Obukhov theory which predicts the intermittency boundary well in $Re-F_h$ space for (wall-normal) stratified plane Couette flow, as shown in figure 18 of \cite{Deusebio:2015}. Although this curve actually delineates the onset of intermittency, as \cite{Deusebio:2015} show in their figure 18, it is also a good estimate of the boundary between flows exhibiting some turbulence and flows which completely relaminarise.} 

Our runs probe as high as $Re=15000$.
{In principle, such large Reynolds numbers
may allow the consideration of 
(still turbulent) flows with }
large enough stratifications to  {potentially} approach 
{the so-called layered anisotropic} stratified turbulence {LAST}
{regime \citep{Brethouwer2007,2016JPO....46.1023F}.  In the LAST regime, there is} a clear separation between the Kolmogorov viscous microscale $l_K$, the Ozmidov scale $l_O$ ({i.e.} the {largest vertical} scale 
{which is not strongly affected by} 
the background stratification) and the typical large {buoyancy layering} scale of the turbulent flow. {This scale separation implies the existence of a highly 
anisotropic strongly stratified yet still turbulent flow for scales
between this buoyancy layering scale and the Ozmidov scale, with essentially isotropic turbulence for scales between $l_O$ and $l_K$.} 

{However, it is not
at all clear that the LAST regime 
can be accessed in flows with wall-forcing,
as all previous numerical investigations
of this regime have (to the authors' knowledge) involved various numerical
body-forcing protocols to ensure 
the maintenance of turbulent motions
in sufficiently strongly stratified flows. However, our intention here }
is not to investigate {any properties of this regime. Rather, we wish to investigate the effect of intermediate (spanwise) stratification on the underlying plane Couette {flow} dynamics, and also whether this  flow geometry can remain turbulent at sufficiently
strong stratifications
to allow the possibility of  transition towards the LAST  regime.} We discuss {the emergent} scale separation and the definitions and emergence of vertical length scales in the following section. 

{Furthermore, we have
conducted a linear stability analysis of this
flow (with $Pr=1$) and as shown by the green shaded region, at sufficiently high $Re$ and small $F_h$, the flow is linearly unstable (provided of course that the unstable vertical and streamwise wavelengths of the instability can `fit' into the chosen computational domain) and so the apparent suppression of the \emph{subcritical} route to turbulence no longer precludes the possibility of turbulence being sustained.
{This linear instability is the analogous instability (for flows with unit Prandtl number) 
to the instability identified by \cite{facchini_2018} at infinite Schmidt number}.
}

\begin{figure}
\begin{centering}
\includegraphics[width=1.\textwidth]{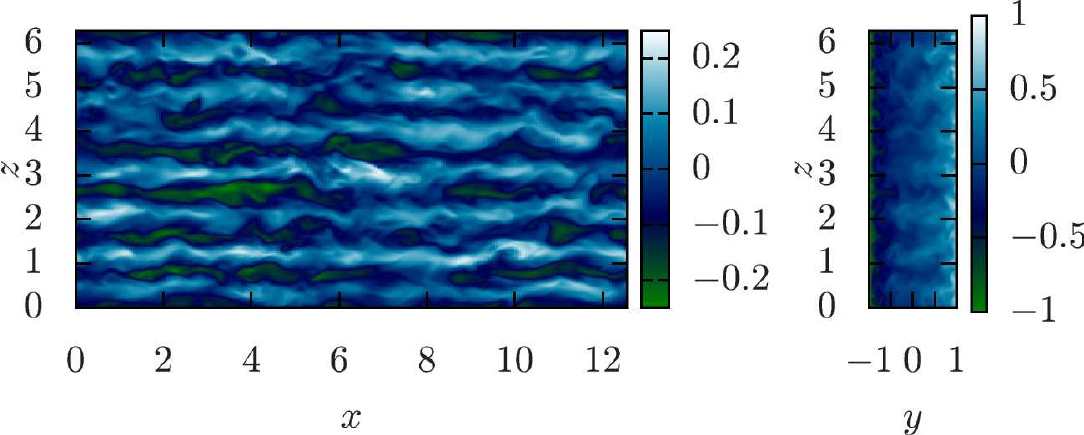}
\caption{\label{fig:U_Re5000}Streamwise velocity, $u,$ for the case $Re=5000$, $F_h^{-2}=0.1$, $Pr=1$, $L_x=4\pi$, $L_z=2\pi$ from set 1 of DNS in table 1. Left {panel shows the flow} at the mid-gap $y=0$ and {the right panel shows the flow} for $x=0.$ Seen are the streaks of high/low speed fluid which have been lifted from near the walls by streamwise rolls, as one expects in plane Couette flow. In these images there is no discernible effect of stratification on the dynamics.}
\end{centering}
\end{figure}

 Figure \ref{fig:U_Re5000} shows some typical flow field snapshots at $Re=5000$ and $F_h^{-2}=0.1,$ which is near to the largest possible stratification possible at this Reynolds number for which {subcritically-triggered} turbulence can be sustained. At first inspection there appears to be little obvious influence of stratification on the typical plane Couette flow dynamics, {as} we {still} observe streaky flow, with streamwise waves propagating along the streaks.

\begin{table}
\begin{center}
\begin{tabular}{ccccccccccccc}
$Re$   &  $ F_h^{-2} $     &    $Re_B$    &      $u_{rms}$     &      $ l_o$&     $l_z$   & $l_S$  &     $ \epsilon $   &    $\chi$  &     $ Fr$      &       $\langle{\mathcal{K}}\rangle_t$    &       $Re_\tau$   &  T \\
\hline
500   & 0.001 & 2192& 0.177 & 16.7 & 2.33 & 2.74 & 0.0044 & 2.1e-5  & 4.37 & 0.0195 & 36.5 & 862\\
600   & 0.002 & 1196& 0.164 & 9.44 & 2.34 & 2.42 &0.0040 & 3.5e-5  & 3.30 & 0.0169 & 41.9 & 852\\
700   & 0.002 & 1316& 0.157 & 9.17 & 2.18 & 2.13 &0.0038 & 3.4e-5 & 3.43 & 0.0159 & 46.9 & 887\\
1000  & 0.005 & 740 & 0.138 & 4.58 & 1.53 & 1.59 & 0.0037 & 6.9e-5 & 2.75 & 0.0136 & 63.0 & 378\\
2000 & 0.005 & 1201& 0.121&  4.13 & 1.51 & 0.862 &0.0030 & 6.56e-5 & 5.80 &  0.0115 &  116 & 363\\
2000 & 0.01 & 562 & 0.118 & 2.36 &  1.37 & 0.893 & 0.00281 & 1.04e-4 & 2.01 & 0.0107 & 112& 356 \\
3000 & 0.01 &749 & 0.108 & 2.24 &  1.29 & 0.629 &0.00250 & 9.78e-5 & 2.15 & 0.0097 & 159 & 292 \\
3000 & 0.05 & 107 & 0.102 &  0.567 &  0.952 & 0.741 & 0.00179 & 1.35e-4 & 0.774 & 0.0072 & 135 & 379 \\
5000&  0.1 & 72.9 & 0.096 & 0.215 & 0.729 & 0.488 &0.0015 & 1.30e-4 & 0.498 & 0.0062&  205 & 453 \\
\hline
5000 &  0.005 & 2445 &  0.113 &  2.63  &  1.375 & 0.385 & 0.0024 &  4.35e-5  &  2.71  & 0.0099 &  260 &  424  \\
5000 &  0.01 &  1135 & 0.105 & 1.51 &   1.122  & 0.398 & 0.0022  & 7.27e-05 & 2.06  & 0.00879  & 250& 452  \\
5000 & 0.0175 & 619 & 0.104 & 0.970 & 0.985 & 0.4032 &0.0022 & 9.97e-05 &  1.51 & 0.0085 & 248 & 586 \\
5000 & 0.025 & 406  & 0.102 &  0.717&  0.995& 0.416 & 0.0020 &  1.15e-4&  1.24 &   0.0079 & 241  &482 \\
5000 & 0.0375 & 262  & 0.102 &  0.521&  0.866& 0.426 & 0.0019 &  1.28e-4&  0.98 &   0.0077 & 237  &621 \\
5000 & 0.05 & 176 & 0.097 & 0.397 & 0.859 & 0.442 & 0.0018 & 1.36e-4 & 0.837 & 0.00696&  225 & 540 \\
5000 & 0.075 & 102 & 0.095 & 0.273 & 0.770 & 0.474 & 0.0015 & 1.29e-4 & 0.617 & 0.00628&  211 & 730 \\
5000 & 0.1 & 63 & 0.093 & 0.200 & 0.672 & 0.541 & 0.0013 &9.73e-5 & 0.463 & 0.00559 & 189 & 666\\

10000&  0.2 &   59.1 & 0.0813 &  0.115 & 0.467 & 0.273 &  0.0012 & 1.27e-4 & 0.40 & 0.00461 &  366 &  450\\
15000 & 0.4  &  26.8 & 0.0759 & 0.053  & 0.319 & 0.234 & 7.14e-4 & 7.02e-05 & 0.20 & 0.00349  &427  & 239 \\
\hline
10000 & 0.2 & 64.0 & 0.0852 & 0.119 &0.540 & 0.263 & 0.0013 & 1.25e-4 & 0.39 & 0.0049 & 380 & 566 \\
10000 & 0.7 & 4.6 & 0.0804 & 0.022 & 0.311 & 0.53 & 3.22e-4 & 2.33e-5 & 0.06 & 0.0041 & 188 & 2500 \\
\end{tabular}
\end{center}
\caption{\label{tab:DNS}Some DNS diagnostics. Top group have $L_x=4\pi,$ $L_z=2\pi,$ $N_x=256,$ $N_z=256,$ $N_y=129.$ Middle group have $L_x=\pi,$ $L_z=2\pi,$ with $N_x=64,$ apart from $Re=10000$ and 15000 which required $N_x=128,$ $N_z=512$ and $N_y=161.$ Bottom group are the two linearly unstable cases $ F_h^{-2}=0.2$ having $L_x=16.5,$ $L_z=5.8$ and  $ F_h^{-2}=0.7$ having $L_x=12,$ $L_z=4.$  All have $Pr=1.$ Averages in time are taken after the initial transient, the time averaging window is given as $T$ in the table. The time averaging window for the linearly unstable case $Re=10000,\, F_h^{-2}=0.2$ is taken once sustained wall-turbulence is reached for comparison to the linearly stable values. $Re_B=\epsilon Re F_h^{2},$  $Fr = \epsilon F_h/(u^2_{rms}),$ $l_O=\sqrt{\epsilon F_h^{3}},$ $Re_\tau=L_y u_\tau/\nu,$ $u_{rms}^2=\langle u'^2 \rangle_V $ {with} $l_z$ computed as discussed in 
section \ref{sec:layers}. }
\end{table}
\subsection{Layers}\label{sec:layers}

\begin{figure}
\begin{centering}
\includegraphics[width=\textwidth]{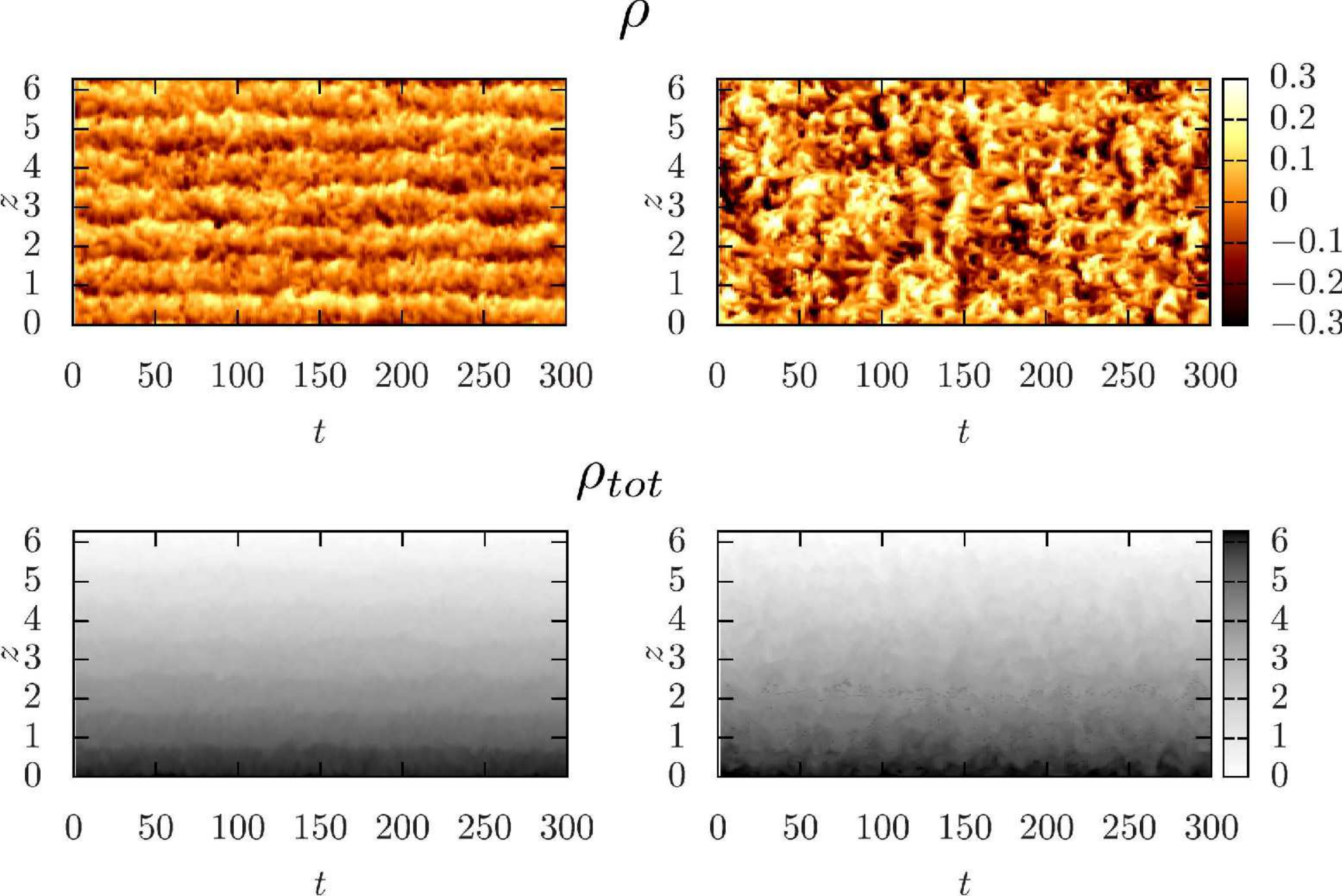}
\caption{\label{fig:TH_ZT} Spatiotemporal diagram in $(z,t)$ for profiles of density for $x=L_x/2$ and two $y$ locations: near wall $y=-0.98$ (left panel) and mid gap $y=0$ (right panel), for the case $Re=5000,$ $F_h^{-2}=0.1,$ $Pr=1,$ $L_x=4\pi,$ $L_z=2\pi$ from set 1 of DNS in table 1. The top panels show the perturbation density $\rho$ in red/yellow, while the bottom panels show the total density $\rho_{tot}=\rho-z$ in grey scale. Note that layers are visible near the wall but there is only well-mixed fluid in the mid-channel.}
\end{centering}
\end{figure}

Despite the apparent weak dependence of the flow on the stratification, some differences can be observed. Figure \ref{fig:TH_ZT} shows spatiotemporal diagrams of the {perturbation} density $\rho$ {(top panels)} and $\rho_{tot}$ {(bottom panels)} in $(z,t)$-planes at fixed $x$ and $y$ for the case $Re=5000,$ $F_h^{-2}=0.1,\, L_x=4\pi,\, L_z=2\pi$ . We show $x=\pi$ and two $y$ locations: $y=-0.98,$ i.e. near-wall; and $y=0,$ mid-gap. The near-wall profiles clearly show the spontaneous formation of layers of alternately stronger and weaker stratification which are robust in time. By contrast, the  {flow in the} centre of the channel is uniformly stratified with broadly isotropic fluctuations.

\begin{figure}
\begin{centering}
\includegraphics[width=0.9\textwidth]{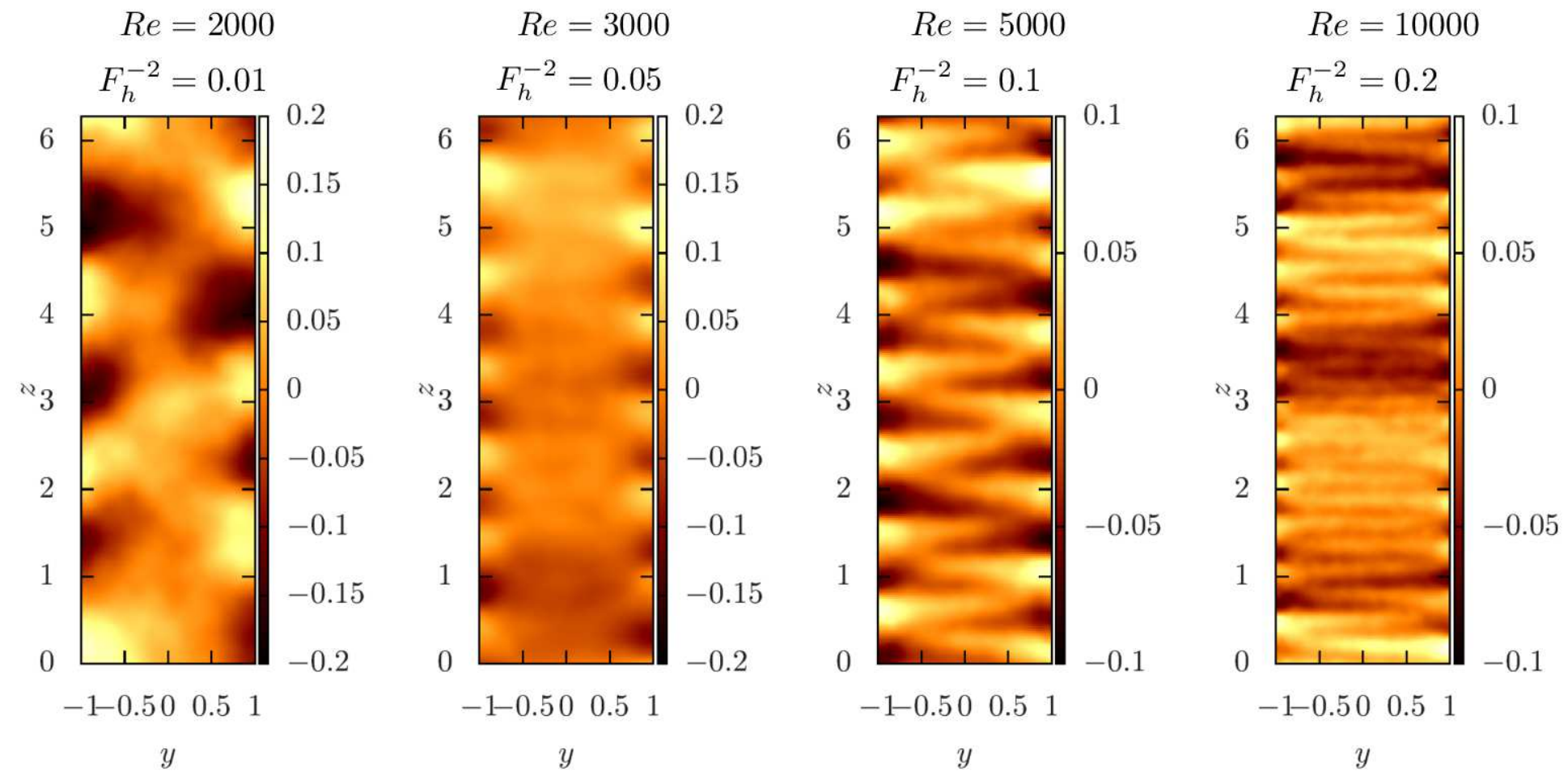}
\caption{\label{fig:rho_YZ} Streamwise- and time-averaged perturbation density $\langle  \rho \rangle_{x,t}$ for 4 cases from set 1 of DNS in table 1 as labelled. Notice the layers near the walls which are offset from each other across the gap.}
\end{centering}
\end{figure}

 Given the robustness in time of the near-wall structures, in figure \ref{fig:rho_YZ} we plot the time- and streamwise- averaged densities in a $(y,z)$-plane i.e. $\langle {\rho}\rangle_{xt}$ for this case and several others from table \ref{tab:DNS}. As $Re$ and $F_h^{-2}$ are increased, coherent layers are observed, offset across the gap, and with decreasing vertical scale.  Associated with this layering is a coherent pattern in the mean flows. {Relatively} large scale flattened streamwise vortices develop with {enhanced} density gradients located between them, near the walls. 
 
\begin{figure}
\begin{centering}
 \includegraphics[width=\textwidth]{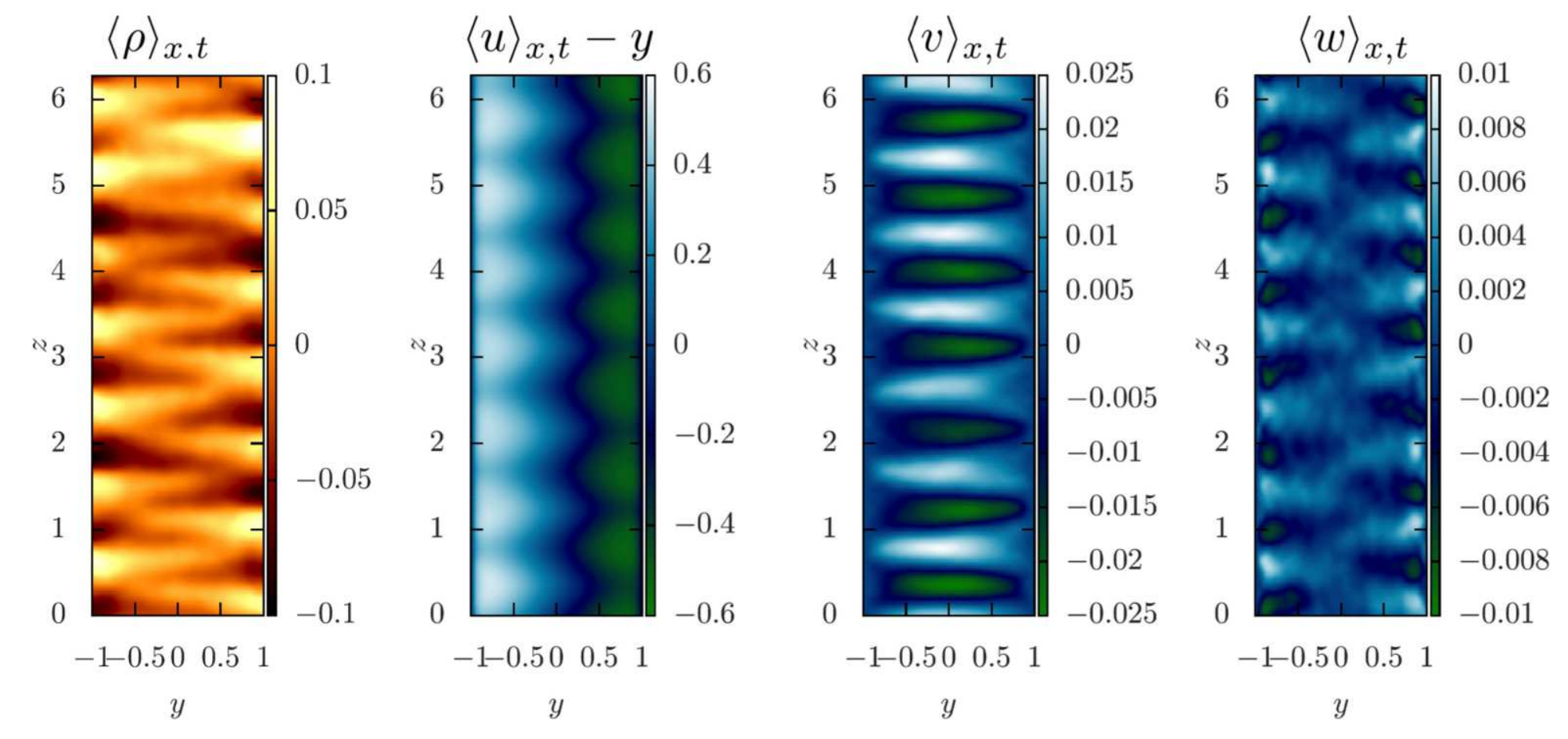}
\caption{\label{fig:vel_mean} Streamwise- and time-averaged total velocities $\langle \bm{u} \rangle_{xt}-y\hx$ for $Re=5000$ $F_h^{-2}=0.1$, $Pr=1$, $L_x=4\pi$, $L_z=2\pi$ from set 1 of DNS in table 1. Streamwise vortices emerge as a coherent structure coupled to the density layering.}
\end{centering}
\end{figure}

{Considering again} the case $Re=5000,$ $F_h^{-2}=0.1,$ $Pr=1$ and $L_x=4\pi$ and $L_z=2\pi,$ figure \ref{fig:vel_mean} shows the streamwise- and time-averaged velocity components of the mean flow. {The apparent} vortical structure {leads to} {vertical motions and hence} buoyancy fluxes localised near the walls and an associated `zig-zag' pattern in the streamwise velocity perturbation. This flow structure bears some resemblance to the exact coherent structures discussed in LCK as alternating spanwise velocities {once again} redistribute the background shear. {A difference} in this case {is that} the wall confinement causes the streamlines to form closed loops rather than penetrate through the periodic boundary as in LCK.
 
 {There is also some similarity to the linear modes discovered in this system by  \cite{facchini_2018} insofar as 
 {there are}
 density perturbations concentrated near the walls. However the length scales and driving mechanisms for these near-wall perturbations are completely different as can be seen in section \ref{sec:instab},
 {and so it may well be that the similarity is merely coincidental.}
  At the largest Reynolds number and stratification considered there is some indication of an additional modulation to this layering pattern, the right most panel of figure \ref{fig:rho_YZ} showing an approximately mode 3 structure on top of the much finer near-wall layers.} We now seek to clarify the underlying mechanism for forming this persistent large-scale flow, and also to investigate how it influences the overall flow dynamics. 
 
\begin{figure}
\begin{centering}
\includegraphics[width=\textwidth]{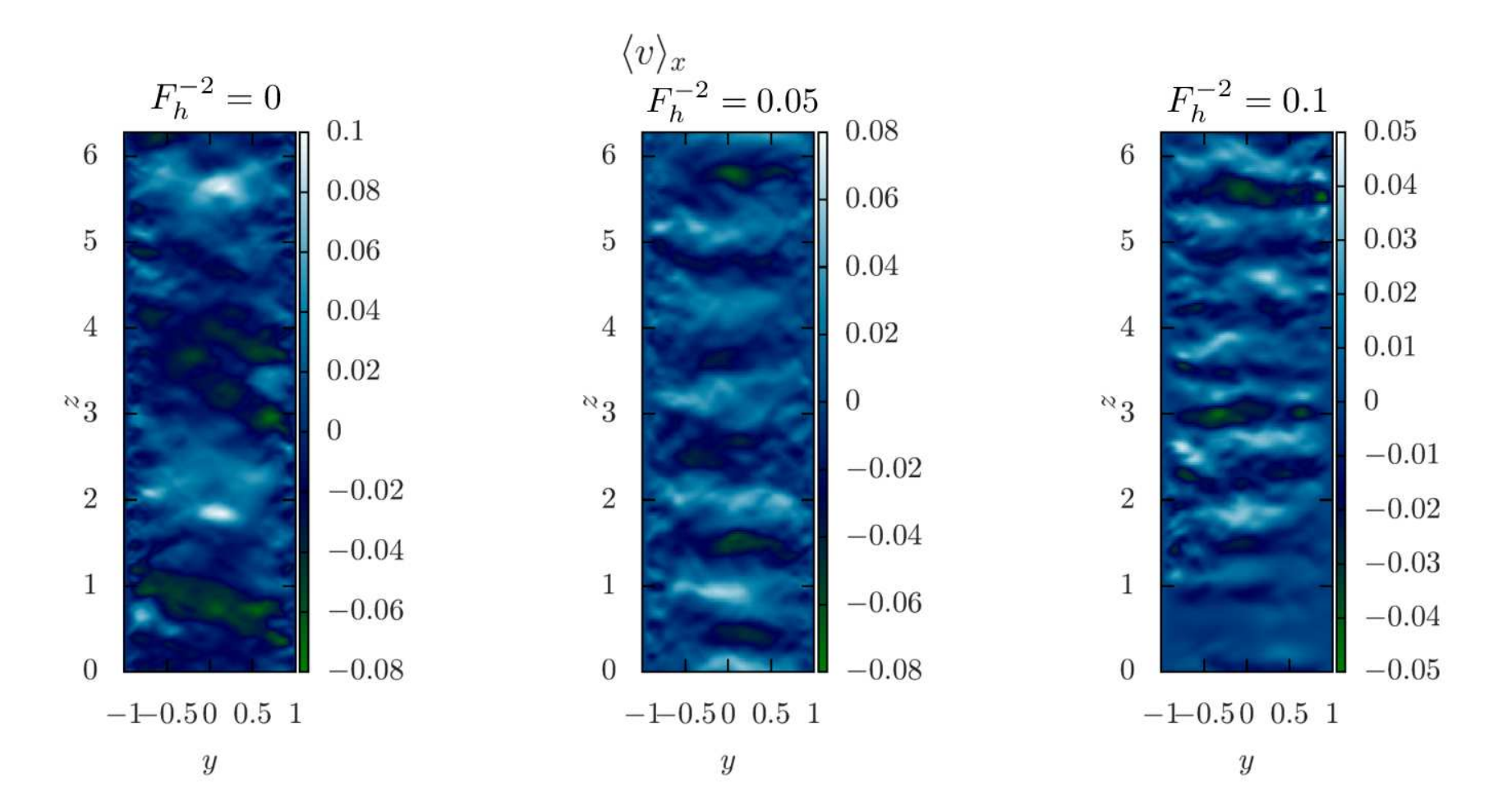}
\caption{\label{fig:vbar} Streamwise-averaged wall-normal velocity $\langle v \rangle_{x}$ for $Re=5000$, $Pr=1$, $L_x=\pi$, $L_z=2\pi$ for $F_h^{-2}=0,$ $F_h^{-2}=0.05$ and $F_h^{-2}=0.1$ from DNS in table 1. The large-scale flow observed in unstratified plane Couette flow is shown in the leftmost panel, with the middle and right {panels} showing how stratification confines the secondary flow to a shallower vertical scale, decreasing with {decreasing} $F_h$. }
\end{centering}
\end{figure}

Having established that this large-scale structure is streamwise invariant in the relatively long streamwise domain $L_x=4\pi,$ i.e. in the averages of figure \ref{fig:vel_mean}, we study {flows with} higher $Re$ and fixed $Re=5000$ with variable $F_h^{-2}$ in a shortened domain $L_x=\pi$ to reduce the computational expense. These are shown in the second set of results in table \ref{tab:DNS}. Note that even the least energetic of these cases has $Re_\tau\approx190$ corresponding to $L_x^+\approx 600$ in viscous units, which should be comfortably above the minimal distance \citep{jimenez_1991}.  

Insight into {the properties of these large-scale structures} is gained by considering the limit $F_h\to \infty$. In unstratified plane Couette flow, streamwise-invariant large scale structures are known to exist at larger Reynolds number as weak secondary flows \citep{Papavassiliou_1997,toh_itano_2005,Tsukahara_2006}. Figure \ref{fig:vbar} shows snapshots of streamwise-averaged wall-normal velocity $\langle v \rangle_x$ for $F_h^{-2}=0,\,0.05$ and 0.1 for $Re=5000,$ which shows this large-scale structure in the unstratified case and the reduction in the vertical length scale of this structure as {$F_h$ decreases}.  These secondary streamwise rolls have been understood as essentially a large scale condensate of an inverse cascade of quasi-two-dimensional streamwise vorticity. These structures are mostly inviscid, except near the walls and therefore require only {relatively} small energy flux to maintain them. A self-sustenance between the Reynolds stresses of the turbulence and the large-scale secondary flow allow the structure to persist in space and time. 

Due to their large scales, in the vertical in particular,  these are the first coherent structures to feel the effect of stratification. As $F_h$ {is decreased}, this large-scale flow becomes constrained in the vertical, $z,$ direction by the {vertical} buoyancy scale $l_z.$  The buoyancy length scale is commonly observed to scale as $l_z\sim U/N$ (where $U$ is a typical horizontal velocity scale) which has been predicted by scaling arguments of the governing equations \cite{Billant:2001cs} and also by linear instabilities (\cite{Billant:2001cs}, LCK). Here we also observe {this characteristic} $U/N$ scaling as shown in figure \ref{fig_scale}, where we have estimated $l_z$ using a wavenumber centroid method on the spectra of $\langle{v}\rangle_{x,y,t}$ similar to {the approach described in} LCK, and chosen $U=v_{rms}.$

\subsection{Relaminarisation}

\begin{figure}
\begin{centering}
\includegraphics[width=\textwidth]{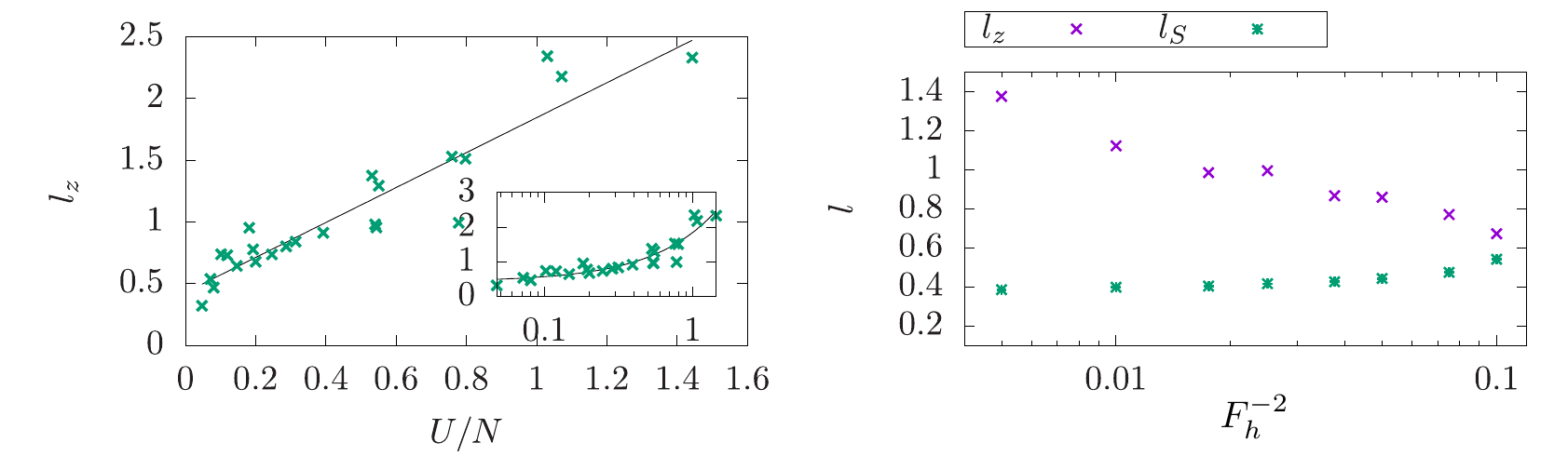}
\caption{\label{fig_scale} Left: Estimate of the vertical length scale of the mean streamwise roll, computed by a centroid of the Fourier transform of mean wall-normal velocity, plotted against an estimate of $U/N$ where $U\equiv v_{rms}$ and $N=F_h^{-1}.$ A linear fit is shown where $l_z= 1.44U/N + 0.39$ and data is given from all cases in table \ref{tab:DNS}. Inset shows the same plotted with $U/N$ on a log axis to better show the cluster of points for $U/N<0.4.$ Right: Vertical length scales against $F_h^{-2}$ for the case of fixed $Re=5000$ from the second set of DNS in table \ref{tab:DNS}. $l_z$ is given as in the left plot, along with $l_S=100\nu/u_\tau.$ Decreasing $F_h$ beyond the values given here results in relaminarisation {of this subcritically-triggered turbulent flow, which may be} interpreted as {occurring due to} the intersection of the buoyancy scale with the streak spacing and subsequent disruption of the SSP/VWI mechanism by stratification. }
\end{centering}
\end{figure}

This interpretation of the layering and influence of the buoyancy scale on this system allows for further interpretation of the relaminarisation boundary. The next largest spanwise length scale in pCf dynamics is the inner streak spacing. This spacing has been established as $l_S=100\nu/u_\tau$ (see e.g. \cite{kline_1967,kim_1987,Hamilton:1995}). By examining the case of fixed $Re=5000$ and varying only $F_h$ we plot in figure \ref{fig_scale} the buoyancy length scale $l_z$ and the streak-spacing $l_S.$ As $F_h$ is decreased the two length scales become closer in value, and their intersection represents the relaminarisation point with respect to $F_h$ at this $Re.$ 

\begin{figure}
\begin{centering}
\includegraphics[width=\textwidth]{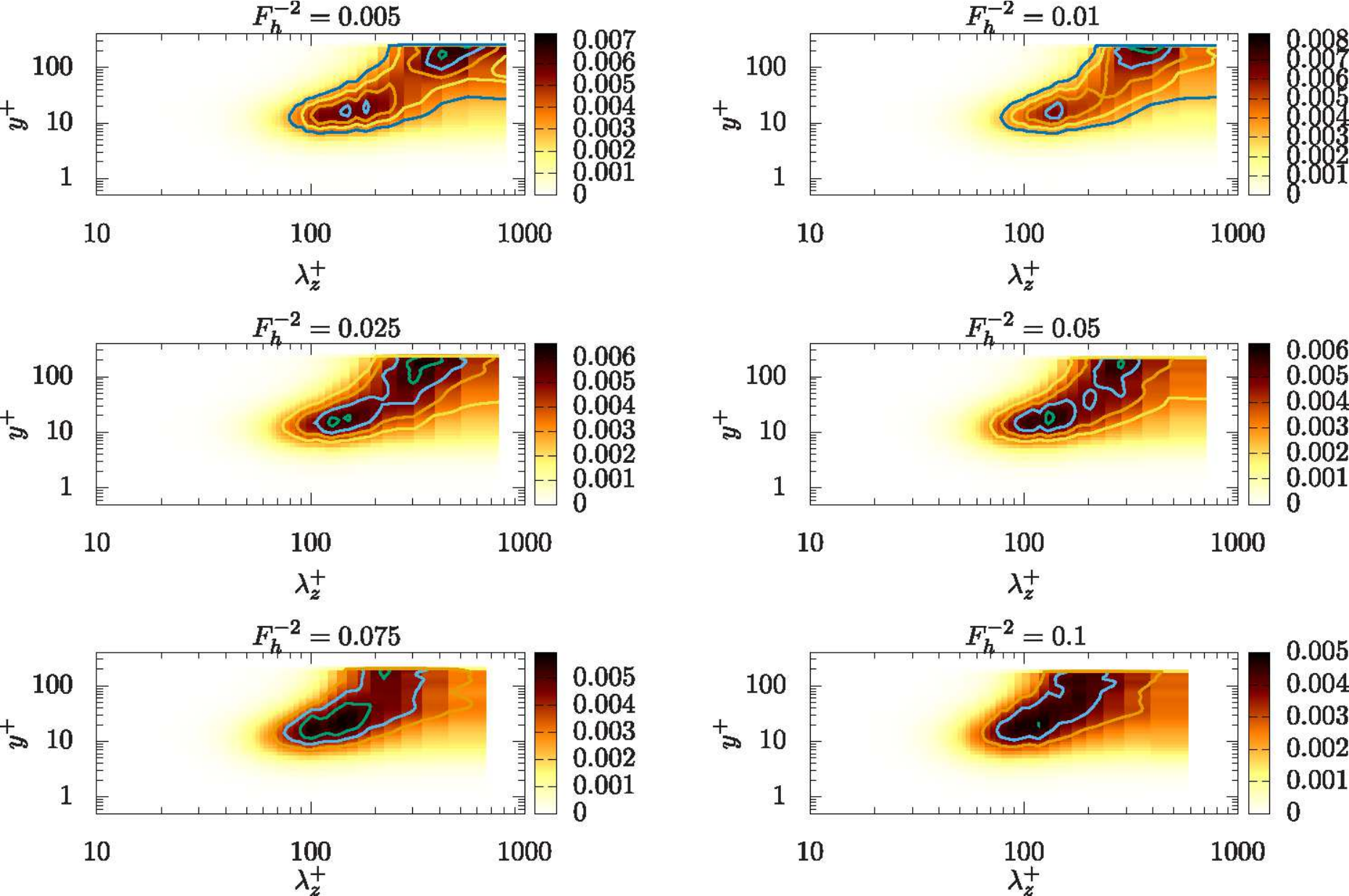}
\caption{\label{fig_spec} Spanwise spectra of $U^+$ plotted against $y^+,$ the distance from the wall in viscous units. The spectra are constructed from $k_x=2\pi/L_x$ i.e. the largest streamwise length scale which is optimal for picking out the signature of the near-wall regeneration cycle at $\lambda^+_z=100$ and the buoyancy scale ($\lambda^+_z\approx400$ at $F_h^{-2}=0.005$). Various values of $F_h$ from the DNS of table \ref{tab:DNS} are shown, decreasing from top left to bottom right. As $F_h$ decreases the near-wall peak stays fixed but the buoyancy scale shifts towards it i.e. decreasing $\lambda^+_z\approx$ and $y^+.$ At $F_h^{-2}=0.1$ the peaks intersect and further decreases in $F_h$ result in overlap of these scales and relaminarisation.}
\end{centering}
\end{figure}

To analyse this {scale}
convergence 
more carefully, we plot in figure \ref{fig_spec} the spanwise spectra with $y$ of the normalised mean streamwise velocity $\hat{U}^+=\langle{\hat{ u}}\rangle_t/u_\tau$ (where $\hat{.}$ denotes the Fourier transform) for streamwise wavenumber $k_x=2\pi/L_x.$ This choice is made to pick out the signature of the largest scale streamwise mean flow and the near-wall fluctuations. {For relatively large $F_h$,} there are two peaks, the inner streak scale corresponding to $l_S=100\nu/u_\tau,$ i.e. at $\lambda_z^+=100$ and $y^+\sim 10$ and the large scale secondary flow further from the wall (in this measure) and with larger spanwise wavelength. As 
{$F_h$ is decreased,} the spanwise buoyancy scale reduces and begins to penetrate towards the wall,
{such that for $F_h=\sqrt{10}$}
the two peaks have essentially merged. Further 
reducing 
{$F_h$}
{causes} 
the buoyancy scale {to envelop completely and overlap}
the inner streak scale. In other words, the small-scale streaks become strongly influenced by the buoyancy scale so that the SSP/VWI mechanism is disrupted and turbulence is unable to maintain itself.

{The discussion of the influence of stratification on SSP/VWI dynamics is developed
in detail in \cite{olvera_2017} and \cite{Deguchi_2017}. Although both these studies focus on wall-normal stratification, the leading effect identified there is the same in spanwise stratification: the presence of stratification in either the wall-normal or spanwise direction inhibits the streamwise rolls which underpin SSP/VWI (e.g. equations (3.12) and (3.13) of \cite{olvera_2017}). This is because in both cases the rolls of the streaks-rolls-waves sustaining cycle are most strongly penalised by the potential energy burden of overcoming gravity. By weakening the rolls, stratification directly reduces the lift-up effect so that the streaks are smaller amplitude and, at some point, may not support instabilities to re-energise the rolls.

The ultimate suppression mechanism is, however, qualitatively different from the essentially wall-dominated mechanism for (wall-normal) stratified pCf discussed in detail in \cite{Deusebio:2015}. There, the key mechanism, as previously identified by \cite{Flores_2010}, is the suppression by stratification of vertical momentum transport essential to the maintenance of turbulence from the near-wall boundary layers, as quantified by the magnitude of the so-called Monin-Obukhov length in wall units. Here, in the spanwise stratification case, suppression comes  through the disruption of the SSP/VWI mechanism near the wall by the  large scale flow. Figure 7(right) indicates that this disruption occurs when  the larger buoyancy scale $l_z$ approaches the streak scaling $l_S$.}

{Also, in} contrast to the triply-periodic body-forced case with horizontal shear {discussed} in LCK, the layers are a relatively simple modification of an existing secondary flow, and not the result of new instabilities or nonlinear exact coherent structures. As indicated above, the stratified version of the self-sustaining process of pCf (i.e. SSP/VWI) still persists at scales below $l_z$ and turbulence is destroyed at larger stratifications. Care is required, however, when interpreting the hierarchy of scales present at small $F_h.$ \cite{Brethouwer2007} detail conditions for `layered anisotropic stratified turbulence' (LAST) to be observed based on a separation of scales such that  $\eta\ll l_O<l_z \ll  l_h$ where $l_O=\sqrt{\epsilon F_h^{3}}$ is the Ozmidov scale, the largest {vertical} scale {largely} unaffected by stratification,  $\eta=(\nu^{3}/\epsilon)^{1/4}$ is the Kolmogorov microscale {and $l_h$ is some appropriate horizontal scale}.  This ordering of scales indicates that an established inertial {dynamic} range of {essentially} isotropic scales must exist below $l_O$ with the LAST regime operating between $l_z$ and $l_O$.

 In table \ref{tab:DNS} a standard estimate of $l_O$ is given. It shows that for the largest stratifications $l_O<l_S,$ from which we might incorrectly infer that the streaks are strongly affected by stratification. This arises from the false assumption that the flow is isotropic below $l_O.$ {In this flow, even when unstratified, the inherent streamwise-spanwise} anisotropy of the SSP/VWI configuration results in weaker fluctuations of {spanwise (or vertical)} velocity relative to the streamwise {velocity fluctuations}, and so for a given level of dissipation and stratification, the vertical velocity is less confined than in a similarly energetic isotropic flow. For this reason we argue that the pertinent vertical length scale of interest is the buoyancy scale $l_z$ as discussed above and the {classical} interpretation of the Ozmidov scale as defined here should be made with care. 

{It is conceivable that with} increased computational resources this system {may possibly} continue into an equivalent LAST regime with larger $Re$ and smaller $F_h,$ and we conjecture that the large scale coherent structures discussed here will persist and the scaling $l_z\sim U/N$ will become clearer. In principle a region in parameter space with $l_z>l'_O>l_S>\eta$ should be observed, where $l'_O$ is a suitably redefined Ozmidov scale for this inherently anisotropic case (i.e. the true scale below which stratification has no influence).  As the relaminarisation boundary is approached we assume the three-way balance $l_z\sim l_S \sim l'_O$ will be attained.  Importantly, we stress that as in LCK, we have shown yet another example of stratification influencing the flow and spontaneously producing layered structures outside of this strict asymptotic regime, driven by an altogether different mechanism to LCK. 


{\section{Linear instability {and supercritical transition}}\label{sec:instab}
{Heretofore, in our discussion of the DNS}
we have neglected the influence of the stratified linear instability which {can} appear in this system, as described in \cite{facchini_2018}. 
{We have only considered flows where}
the dynamics below $l_z$ are similar to the SSP/VWI subcritically-triggered turbulence observed without stratification.
{This can be justified by considering the results of} an {independent} linear stability analysis,
{leading to the estimate}
of the neutral curve on the $(F_h,\, Re)$ plane for $Pr=1$ 
{plotted on} 
figure \ref{fig_TKE}. This curve represents the boundary above which ({shaded in green}) 
it is possible to find wavenumber combinations $k_x$ and $k_z$ for which the basic flow is unstable. Note that for {all} the cases discussed 
{above, we have deliberately}
chosen domains which do not support the linear instability. In addition to this, {it is clear that} only for $Re>5000$ is the relaminarisation boundary disrupted by the linear instability. 

At larger $Re$ and smaller $F_h$ 
{it is necessary to consider} the influence of the linear instability on the relaminarisation and the sustenance of turbulence.
In order to investigate the {potentially supercritically-triggered turbulent} dynamics above the projected relaminarisation boundary of the {subcritically-triggered} turbulence,
we have performed a {simulation} at $Re=10000,\, F_h^{-2}=0.7$ in a geometry $L_x=12,\,L_z=4$, denoted with a triangle symbol in figure \ref{fig_TKE}. ({Note that this  simulation is not listed in table \ref{tab:DNS}.)} In this geometry
the fastest growing linear mode 
{has} one wavelength {in the streamwise $x-$direction} and four {wavelengths} in {the vertical} $z-$direction.
The initial exponential growth rate of the linear normal modes is essentially constant and eventually gives way to nonlinear dynamics at finite amplitude {as} shown by snapshots of $u$ and $w$ in figure \ref{fig:Bpt7}. {(See also accompanying movie (Movie1.gif) in the supplementary material.}

Since we are above the 
{relaminarisation boundary for}
{SSP/VWI} {turbulence},
the nonlinear dynamics remains {somewhat} more regular than below the boundary in the sense of being supported by narrower spectra in space and time. The traces of energetic quantities (figure \ref{fig:energy_Re10000}) have longer time scales with larger amplitude fluctuations, with larger $\mathcal{K}$ and smaller $\epsilon$ than the 
{SSP/VWI} turbulence at {the same} $Re.$ The snapshots of the flow show larger-scale internal wave motions, compared to the smaller scale {SSP/VWI turbulence}
{as can be seen through comparison of} figure \ref{fig:Bpt7} ($t=980$ panel) 
{and} figure 
\ref{fig:Bpt2} ($t=1136$ panel). 
\begin{figure}
\begin{centering}
\includegraphics[width=0.45\textwidth]{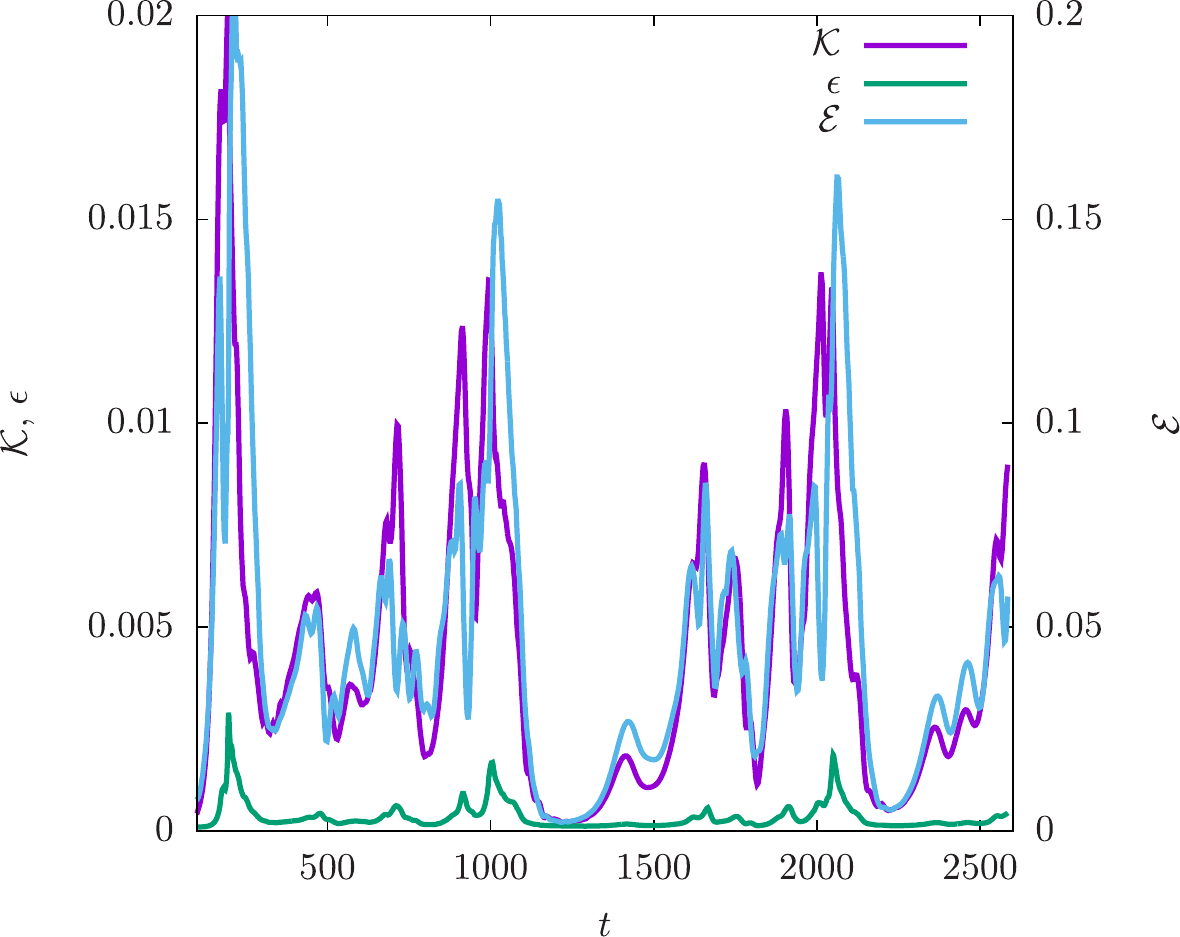}\hspace{10mm}
\includegraphics[width=0.38\textwidth]{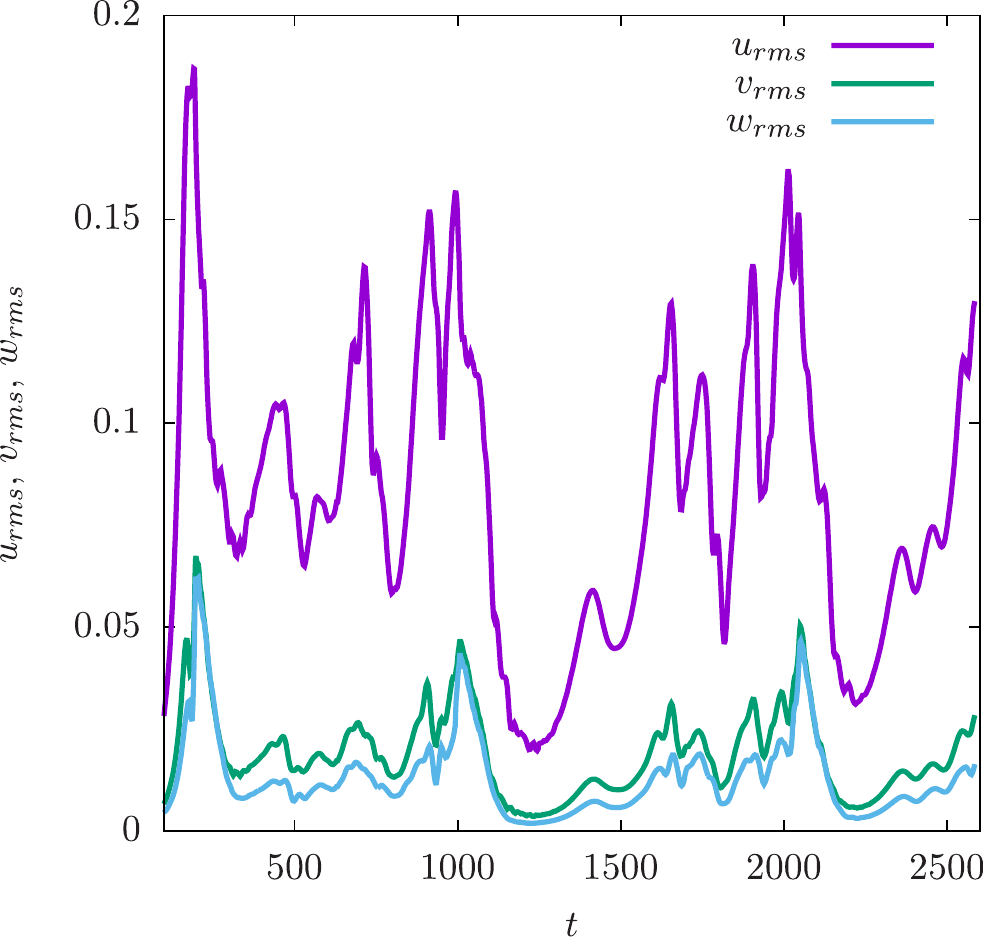}\\
\includegraphics[width=0.45\textwidth]{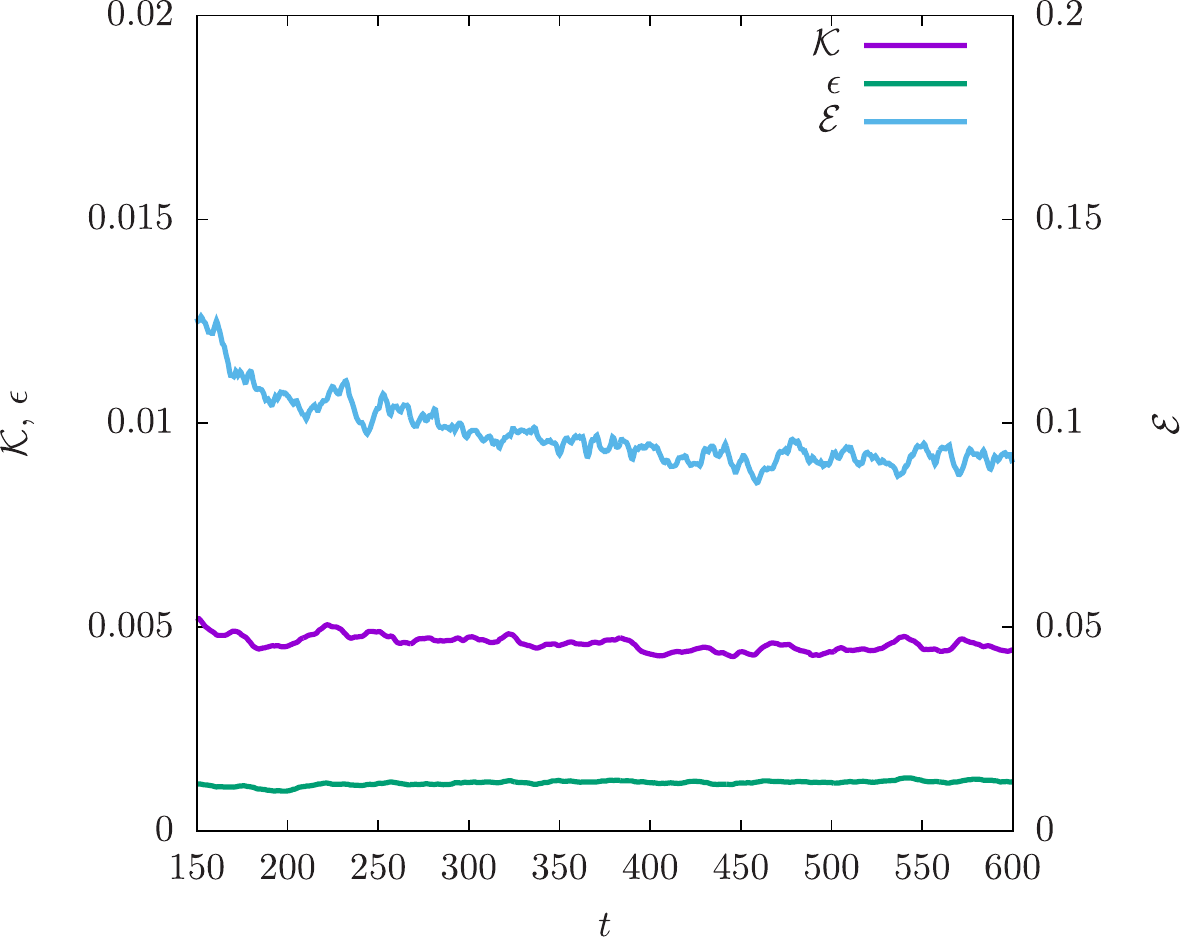}\hspace{10mm}
\includegraphics[width=0.38\textwidth]{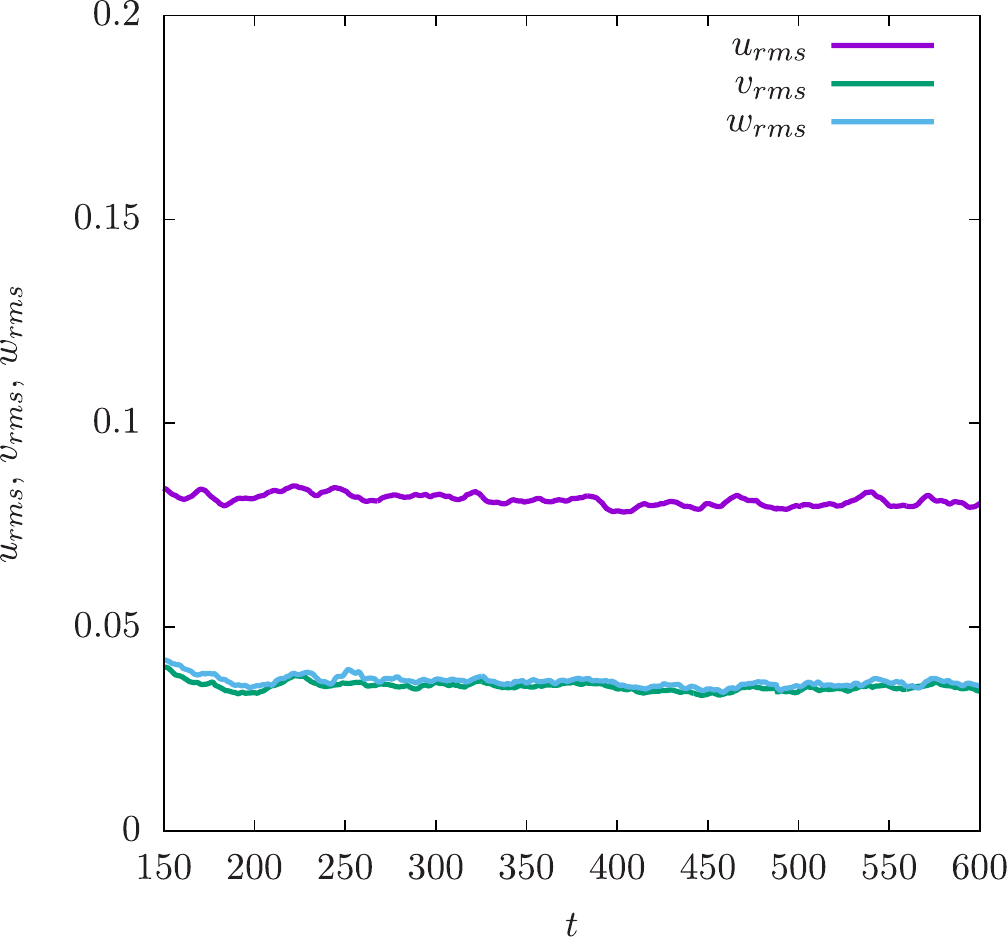}\\
\includegraphics[width=0.45\textwidth]{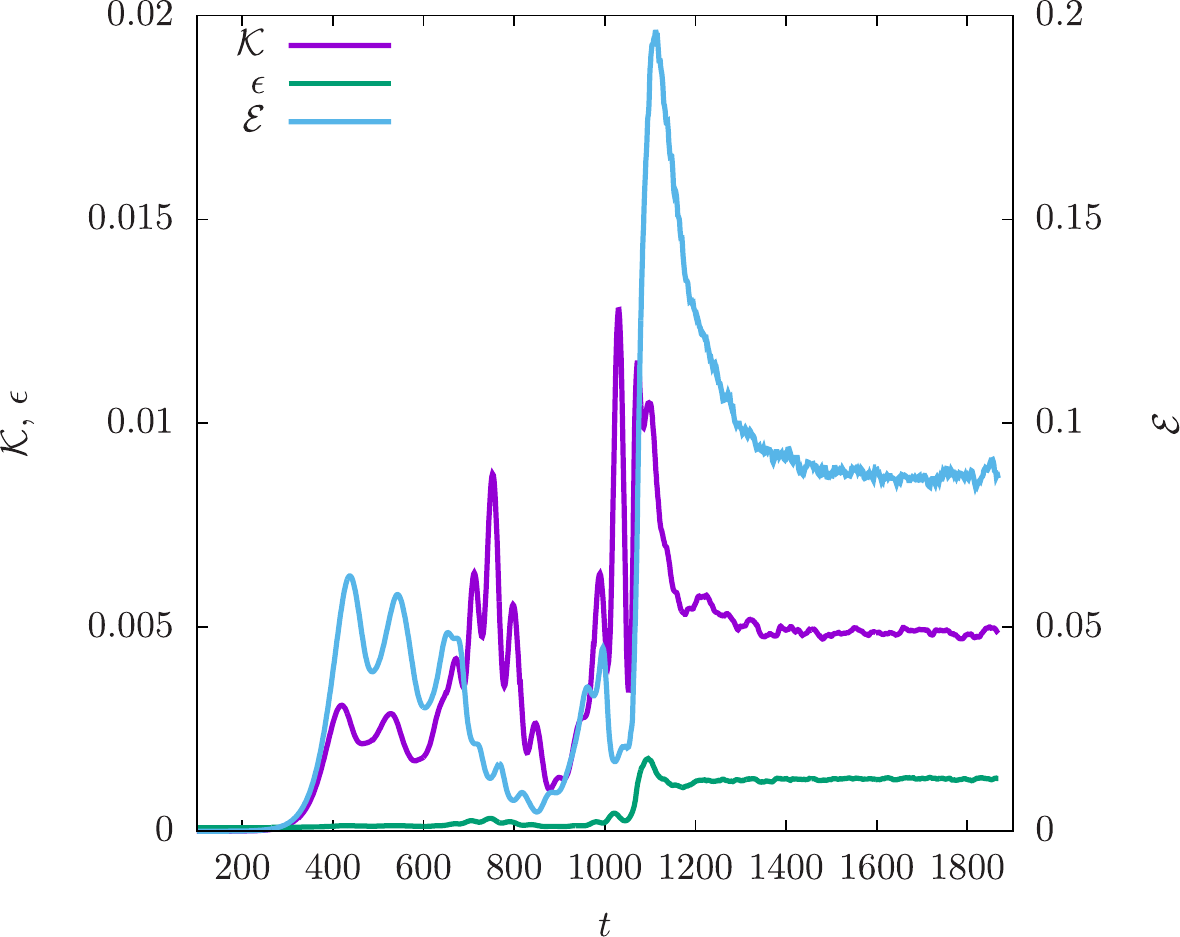}\hspace{10mm}
\includegraphics[width=0.38\textwidth]{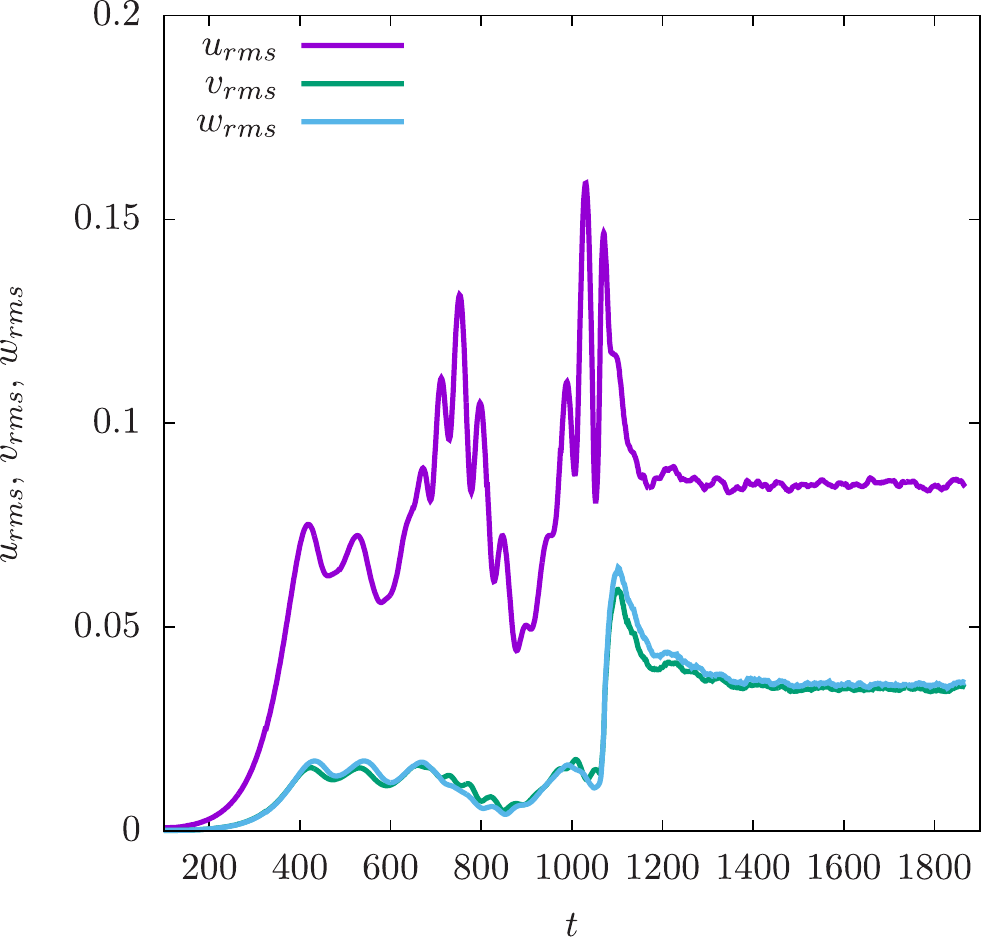}
\caption{\label{fig:energy_Re10000} {Variation with time of:} turbulent kinetic energy $\mathcal{K}$; dissipation $\epsilon$; mixing efficiency $\mathcal{E}=\frac{\chi}{\chi+\epsilon}$ (left) \& r.m.s. velocities (right) for 
{three flows with $Re=10000$ and:}
$F_h^{-2}=0.7$ (top); $F_h^{-2}=0.2$ 
{in a linearly stable geometry with}
($L_x=\pi,\,L_z=2\pi$) {(middle)}; $F_h^{-2}=0.2$ 
{linearly unstable  geometry with} ($L_x=16.5,\,L_z=5.8$) {(bottom)}. Striking are the large fluctuations associated with the nonlinear saturation of the linear instability compared to the SSP/VWI turbulence {below the relaminarisation boundary}
which is the long time attractor for both $F_h^{-2}=0.2$ cases.
{Note the very different time limits for the three simulations.}}
\end{centering}
\end{figure}

There are {also} episodes of more turbulent behaviour where the dissipation rate and mixing efficiency increase and smaller scales are observed in the flow fields. The times of maxima in the turbulent kinetic energy, dissipation and mixing are coincident with local accelerations of the flow and shear-induced overturns 
{associated with}
instability of the internal waves (see the bottom two panels of figure \ref{fig:Bpt7}), {due presumably to}
either wave-wave or wave-mean flow resonance, see \citep{McComas_1977,Grimshaw_1977,sutherland_2001}, and will be a topic of future research. {There is also some suggestion of a
{characteristic} recurrence time scale for the bursting events in the energetics shown in figure \ref{fig:energy_Re10000}, drawing some similarity to the bursts described in LCK ({see in particular} Appendix A.1, which we may expect to recur if the simulation were continued.})\\

At $Re=10000,\, F_h^{-2}=0.2$ it is {actually} 
possible to choose a geometry which 
supports the  {(supercritical)} linear instability
{while still} remaining {below} the boundary for {SSP/VWI turbulence.}
 {This particular flow geometry allows us to consider the competition} 
between the nonlinear dynamics and {different} possible routes to turbulence. 
We perform another DNS at 
{this $Re$ and $F_h$} with $L_x=16.5$ and $L_z=5.8$ (now with $N_x=N_z=512$) and initialised with a large-scale {yet} small-amplitude initial condition. In this geometry we also expect the fastest growing linear mode to have one wavelength in {the} $x-${direction} and four {wavelengths} in {the} $z-${direction}. In this case we see again the 
{initial exponential} growth of the linear normal mode, saturating with the nonlinear internal wave dynamics, {and} spending a significant time there ({as shown in the} top two panels of figure \ref{fig:Bpt2}). 
\begin{figure}
\begin{centering}
\includegraphics[width=0.9\textwidth]{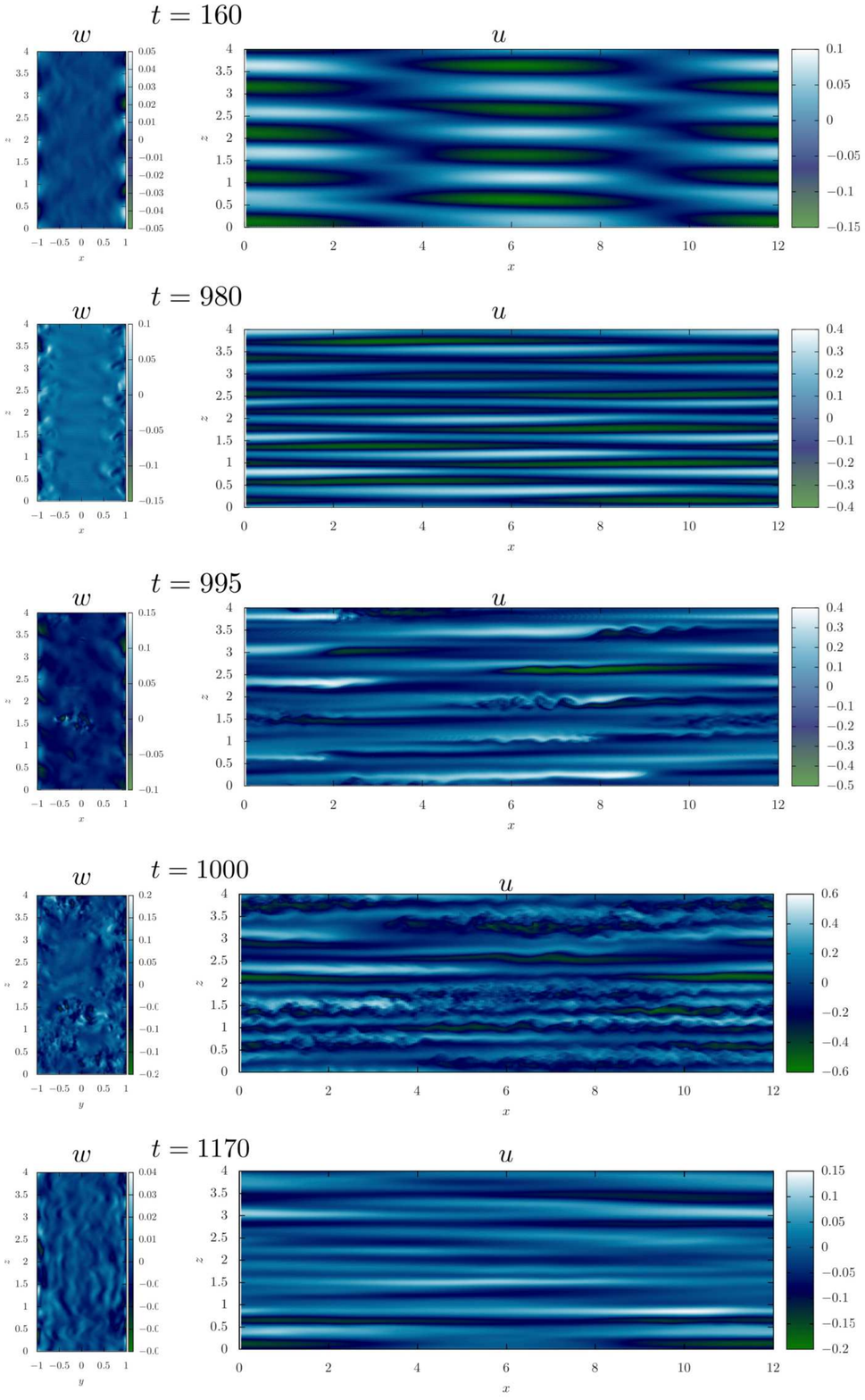}
\caption{\label{fig:Bpt7} Snapshots of $w$ in $(y,z)$ planes (left) and $u$ in $(x,z)$ planes (right) at various times for the linearly unstable $Re=10000,\, F_h^{-2}=0.7$ {flow geometry}:
{a) Early time $t=160$}
shows the signal of linear normal modes;
{b) intermediate time $t=980$ shows the growth of wave-induced shear and smaller wavelengths};
{c) intermediate time $t=995$ shows the development of Kelvin-Helmholtz type overturning;}
d) intermediate time $t=1000$ shows the transient growth of streaky flow;
e) late time $t=1120$ shows the eventual return to the more regular nonlinear wave state. See also accompanying movie (Movie1.gif) in the supplementary material.
}
\end{centering}
\end{figure}

\begin{figure}
\begin{centering}
 \includegraphics[width=\textwidth]{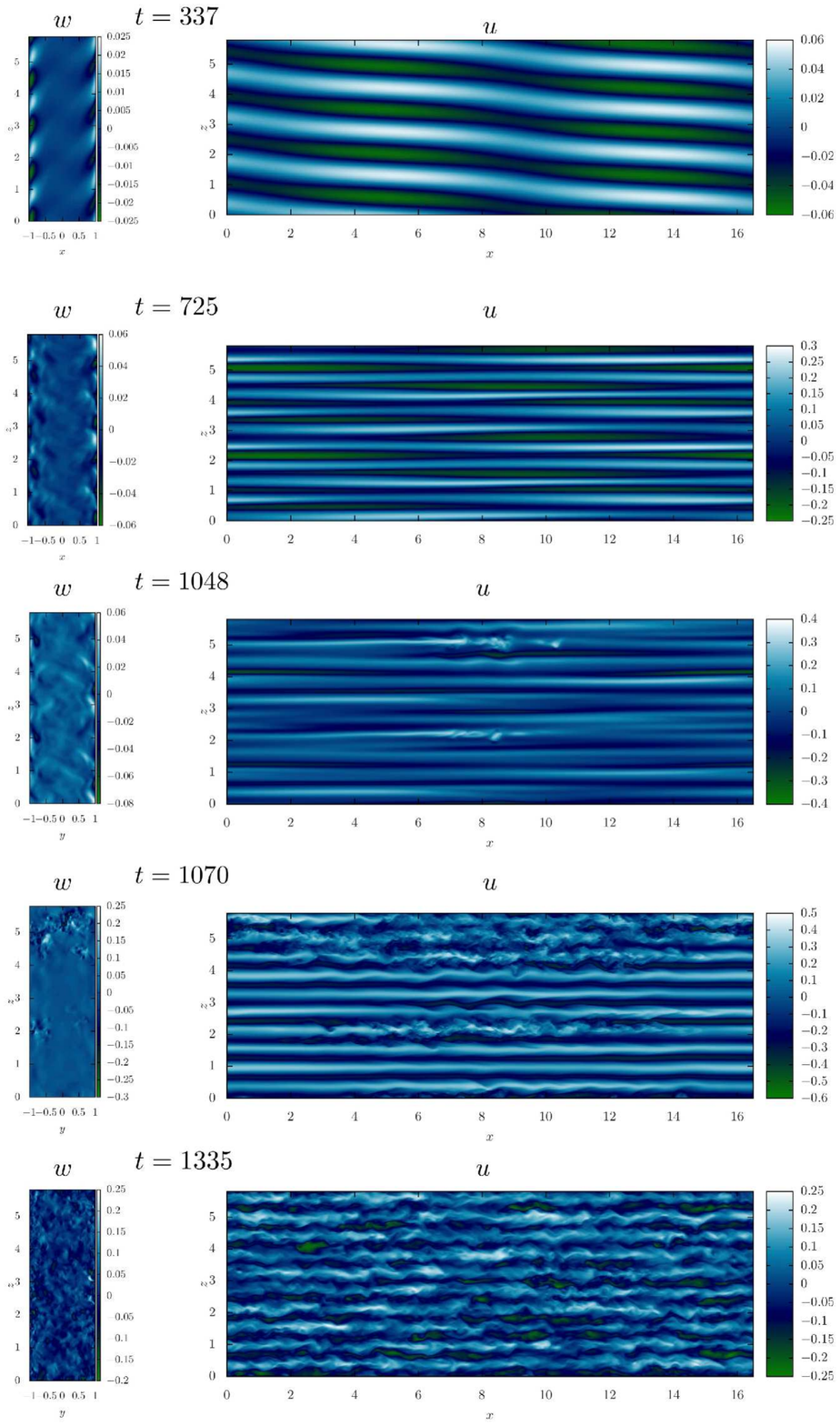}
\caption{\label{fig:Bpt2} Snapshots of $w$ in $(y,z)$ planes (left) and $u$ in $(x,z)$ planes (right) at various times for the linearly unstable $Re=10000,\, F_h^{-2}=0.2$ {flow geometry}:
{a)}
early time {$t=337$} shows the signal of linear normal modes; {b)} intermediate time {$t=725$} {shows}  nonlinear internal waves; {c) intermediate time $t=1048$ shows}  the growth of wave-induced shear and smaller vertical wavelengths; 
{d) intermediate time $t=1070$ shows the ensuing}  Kelvin-Helmholtz type overturning; {e) late time $t=1335$ shows the eventual and} sustained growth of streaky flow indicative of the SSP/VWI attractor. See also accompanying movie (Movie2.gif) in the supplementary material.
}
\end{centering}
\end{figure}
However, this is not the long time attractor.
{Significantly,} the finite amplitudes reached are sufficient to seed the SSP/VWI  {sustained} turbulence which was observed in the {linearly} stable geometry. In other words, 
{apparently,} the linear instability simply provides another route to what appears to be the same attractor at these parameter values. This is borne out by the flow fields in figure 
{\ref{fig:Bpt2}} and the energetics of figure \ref{fig:energy_Re10000} (see also accompanying movie (Movie2.gif) in the supplementary material). We find that when the local accelerations are energetic enough to give rise to overturns (as in the flow with $F_h^{-2}=0.7$) the transfer of energy to small scales initiates the near-wall regeneration mechanism and {sustained SSP/VWI} turbulence is established as the late time attractor (see bottom 3 panels of figure \ref{fig:Bpt2}). The energetics in figure \ref{fig:energy_Re10000} also approach the values {associated with the linearly stable flow, which has undergone subcritical transition} at late times, 
{though there is undoubtedly a very significant (and long-lived) transient adjustment}. It is worth noting that the flow fields after overturning in the $F_h^{-2}=0.7$ case appear to show some transient growth towards an SSP/VWI configuration before stratification suppresses the spanwise streak scale, the small scales decay viscously and the internal wave regime recurs. This can be partially observed in the final frame of figure \ref{fig:Bpt7} and in the r.m.s., energetics and mixing efficiency reaching similar levels {as observed in} the wall-{localised} turbulence case in figure \ref{fig:energy_Re10000}.  { Extrapolating then towards the relaminarisation boundary we expect such transient growth to increase until the boundary is crossed and the streaky flow becomes sustained.}\\

We conclude that below the relaminarisation criterion described in section \ref{sec:layers} the turbulent attractor looks unaffected by the new linear instability,
{although it is undoubtedly important that}
a new linear {instability} mechanism has {been identified} to provide a route for small perturbations to excite turbulence. 
}

\section{Discussion \& conclusions}

The computations performed here have revealed another example of coherent 
layering {of an initially linear density distribution} at moderate stratification. This has more than a passing resemblance to the exact coherent structures isolated in LCK: shear-wise flow advects the streamwise velocity component to create an alternating, or zig-zagging structure. Confinement by the walls results in the shear-wise flow being associated with a streamwise vortex, whereas in the periodic geometry the shear-wise flow is periodic in $y.$ This is analogous to the rotation/advection of the vortex dipole necessary for the zig-zag instability \citep{Billant:2000wg}. Importantly however, in the wall-bounded case the basic flow profile remains linearly stable in the parameter regimes studied here, meaning that the layering structures have a different generating mechanism. (In theory it may be possible to perform a homotopy of an exact coherent structure from a system with a stress-free boundary condition and horizontally shearing base flow which experiences a zig-zag type linear instability and {thus} connect to the no-slip boundary condition considered here.)

The results here bear some similarity to the buoyancy patterns described in \cite{Leclercq:2214839} for the axially stratified Taylor-Couette flow case,  in so far as having sharper gradients confined near the walls and offset across the gap. We have also associated these density perturbations with a large scale vortical flow, however it seems that the underlying mechanisms producing these structures are different again. Here we find the large-scale flow to be the modification of an existing secondary flow in plane Couette {flow}, whereas the suggestion in \citep{Leclercq:2214839} is of a different nonlinear mechanism incorporating spiral modes,
{associated with various linear instabilities of the underlying flow, as also discussed in \cite{Park_2017} and \cite{Park_2018}.} {How these two mechanisms connect as the narrow-gap limit is approached is therefore an interesting open question.}

The other main finding of this work is 
{the extent to which} the laminar-turbulent boundary becomes affected by stratification in this {flow geometry} and in particular how the mechanism responsible for shutting off turbulence as stratification is increased has a fundamentally different interpretation 
{from the flow with}

vertical shear/wall-normal stratification. Here,  the intersection of a buoyancy-dominated large scale and the smaller inner scale of the near-wall regeneration mechanism {is  the key process by which}
relaminarisation 
{occurs.}
This is in contrast to the vertical shearing case where the length scale associated with the flux of momentum into the interior is modified by stratification, 
{as quantified by the magnitude of the Monin-Obukhov length in wall units (see \cite{Flores_2010} and \cite{Deusebio:2015} for more details)}
and found to predict relaminarisation.
{One may also make the observation that in the vertically shearing case buoyancy directly penalises the wall-normal flow responsible for lifting up the wall fluid to form the streaks. On the other hand, in the horizontally shearing orientation this wall-normal flow is indirectly penalised via its coupling with spanwise velocity in the streamwise rolls. Both cases limit the rolls, however the subtle, but crucial, difference regarding the velocity components is the reason for higher buoyancy forces (i.e. $F_h^{-2}$ or $Ri$) at large Reynolds number allowing sustained turbulence in the spanwise stratified orientation.} 

{Two important features of subcritical transition which we have not investigated thoroughly are computational domain size and initial conditions. We conjecture that our relaminarisation boundary may also be considered an `intermittency boundary' and that there will only be some weak dependence of the turbulent fraction with $F_h$ at a given $Re$ (similar to figure 18 in \cite{Deusebio:2015}). Recall, the mechanism is ultimately the suppression of streamwise rolls by buoyancy. 
This can be considered a local, high Reynolds number process, independent of viscosity. We may then conjecture that this relaminarisation mechanism is (largely) independent of box size and therefore we may only expect intermittency when $l_z\approx l_S.$ Likewise we  expect the effect of initial conditions to be modest. We have shown how the turbulent attractor is disrupted by stratification and have not concerned ourselves with the particular pathways to that attractor. A more thorough investigation of these aspects is left for future research.}
 

The results presented here open a number of further research questions and opportunities. It would be of interest if these {inherently nonlinear (i.e. not obviously
and directly connected to a linear formation mechanism) and turbulent} layers are observable in a laboratory experiment similar to those conducted by \cite{facchini_2018}. {We note here that all of the physical and numerical experiments conducted by \cite{facchini_2018}, with the exception of cases where $F_h=\infty,$ are well above the relaminarisation boundary for subcritical turbulence that we have established.} This may open a new avenue for connecting laboratory and numerical experiments involving layers and pattern formation in stratified turbulence. Robustness of these layers to changes in Prandtl number is therefore an obvious consideration; using salt as a stratifying agent increases the Prandtl {number}, or {more precisely the} Schmidt number, to $\approx 700.$ 
{It is at least plausible that}
even sharper gradients of density {might arise in flows} at larger $Pr,$ and 
{it is also conceivable that these layers might}
penetrate further into the interior. 

{Furthermore, the} boundary conditions on $\rho$ are chosen to satisfy a no-flux condition. However, if heat were used as the stratifying agent then 
{it might be expected that there would be} some heat loss through the side-walls. A test case (results not shown) suggests that applying an insulating boundary condition, i.e. setting $ \left.\rho \right|_{y=-L_y} = \left. \rho\right |_{y=L_y}=0$ yields very similar near-wall layers and large scale mean flows, suggesting a certain robustness of the {bulk flow features reported here.}

Our results also highlight an interesting question concerning the role of the orientation of shear in more general environmental and industrial flows. For example, in pipes and ducts the mean flow will 
{generically} have  shear {at some non-trivial angle to the orientation of gravity.} 
As noted previously,  horizontal shear, {particularly at sufficiently high $Re$, is} 
a more effective route to transfer energy from the mean flow to the turbulence \citep{Jacobitz:1998di}
{than purely vertical shear,}
and as we have shown the turbulence suppression mechanisms are quite different for each case. The question then arises 
{as to how these mechanisms compete, and also how}
turbulence is deformed by stratification in 
 more generic situations. Just to mention one example, it has recently been observed that horizontal confinement {in a duct, and
hence the inevitable introduction of spanwise shear} to the Holmboe instability of a principally vertically sheared {(relatively `sharp')} density interface gives rise to a new robust long-lived nonlinear coherent structure in the form of  a `confined Holmboe wave' \citep{Lefauve_2018}. It remains to be seen how similar confinement {might affect turbulent flows, even without the sharp density interface typically required
for the initial occurence of Holmboe-type instabilities.}

Finally,  \cite{Taylor_2016} have shown how stratification can be a useful tool {for studying the} localisation and control of turbulence in  vertically sheared flows. It is widely understood that large scale flows play some role in  pattern formation and {the transient} growth {and/or} decay of localised turbulence \citep{lemoult_aider_wesfreid_2013,Duguet_2013,Couliou_2015}.
{Here we have demonstrated} how relaminarisation {can be} controlled by the buoyancy scale. It is therefore clearly of interest to investigate how the modification of large scale flows discovered here influences the spatiotemporal behaviour of localised turbulence, and to what extent the control method of \cite{Taylor_2016} can be
utilised
to facilitate such an investigation. 
{We intend to report the results of just such an investigation in due course.}\\

\noindent
{\em Acknowledgements}. 
We extend our thanks, for many helpful and enlightening discussions, to Paul Linden, John Taylor, Stuart Dalziel and the rest of the `MUST' team in Cambridge and Bristol. We also offer a particular thanks to Qi Zhou for supplying us with the Monin-Obukhov generated intermittency boundary curve for the wall-normal stratified case. The source code, including linear stability code, is provided at \url{https://bitbucket.org/dan_lucas/spanwise-stratified-diablo} and accompanying data can be found at \url{https://doi.org/10.17863/CAM.37838}. This work is supported by EPSRC Programme Grant EP/K034529/1 entitled `Mathematical Underpinnings of Stratified Turbulence'. The research presented here was initiated when DL was a postdoctoral researcher in DAMTP as part of the MUST programme grant.
\bibliography{papers}
\bibliographystyle{jfm}

\end{document}